\documentclass[sigconf]{acmart}
\usepackage{graphicx}
\usepackage{multirow}
\usepackage{booktabs} 
\usepackage{tabularx}
\usepackage{subfigure}

\usepackage{balance}
\def\BibTeX{{\rm B\kern-.05em{\sc i\kern-.025em b}\kern-.08emT\kern-.1667em\lower.7ex\hbox{E}\kern-.125emX}}
    
\copyrightyear{2019} 
\acmYear{2019} 
\acmConference[CIKM '19]{The 28th ACM International Conference on Information and Knowledge Management}{November 3--7, 2019}{Beijing, China}
\acmBooktitle{The 28th ACM International Conference on Information and Knowledge Management (CIKM '19), November 3--7, 2019, Beijing, China}
\acmPrice{15.00}
\acmDOI{10.1145/3357384.3357951}
\acmISBN{978-1-4503-6976-3/19/11}

\begin{document}

\fancyhead{}
  % do not delete this code.

% The "title" command has an optional parameter, allowing the author to define a "short title" to be used in page headers.
\title{Fi-GNN: Modeling Feature Interactions via Graph Neural Networks for CTR Prediction}

% The "author" command and its associated commands are used to define the authors and their affiliations.
% Of note is the shared affiliation of the first two authors, and the "authornote" and "authornotemark" commands
% used to denote shared contribution to the research.

\author{Zekun Li}
\affiliation{Institute of Information Engineering, Chinese Academy of Sciences}
\affiliation{University of Chinese Academy of Sciences}
\email{lizekunlee@gmail.com}
\author{Zeyu Cui}
\affiliation{Institute of Automation, Chinese Academy of Sciences}
\affiliation{University of Chinese Academy of Sciences}
\email{zeyu.cui@nlpr.ia.ac.cn}
\author{Shu Wu}
\affiliation{Institute of Automation and \\Artificial Intelligence Research, Chinese Academy of Sciences}
\email{shu.wu@nlpr.ia.ac.cn}
\author{Xiaoyu Zhang}
\affiliation{Institute of Information Engineering, Chinese Academy of Sciences}
\email{zhangxiaoyu@iie.ac.cn}
\author{Liang Wang}
\affiliation{Institute of Automation, Chinese Academy of Sciences}
\affiliation{University of Chinese Academy of Sciences}
\email{wangliang@nlpr.ia.ac.cn}
\thanks{The first two authors Zekun Li and Zeyu Cui contribute to this work equally.
Shu Wu and Xiaoyu Zhang are both corresponding authors.}

\renewcommand{\shortauthors}{Li and Cui, et al.}

%
% The abstract is a short summary of the work to be presented in the article.
\begin{abstract}
Click-through rate (CTR) prediction is an essential task in web applications such as online advertising and recommender systems, whose features are usually in multi-field form.
The key of this task is to model feature interactions among different feature fields.
Recently proposed deep learning based models follow a general paradigm: raw sparse input multi-filed features are first mapped into dense field embedding vectors, and then simply concatenated together to feed into deep neural networks (DNN) or other specifically designed networks to learn high-order feature interactions. 
%However, the unstructured combination of different feature fields will be a bottleneck, which brings difficulty to model sophisticated interactions among the fields in an enough effective and explicit fashion.
However, the simple \emph{unstructured combination} of feature fields will inevitably limit the capability to model sophisticated interactions among different fields in a sufficiently flexible and explicit fashion.

In this work, we propose to represent the multi-field features in a graph structure intuitively, where each node corresponds to a feature field and different fields can interact through edges.  
%On the graph, it is feasible to model sophisticated interactions among fields in the form of node interactions.  
%To this end, we design a novel model Feature Interaction Graph Neural Networks (Fi-GNN) based on Graph Neural Networks (GNN). 
%It can not only model sophisticated feature interactions in a more effective and explicit fashion, but also provide good model explanations, which are beneficial for CTR prediction, especially recommender systems.
The task of modeling feature interactions can be thus converted to modeling node interactions on the corresponding graph.
To this end, we design a novel model Feature Interaction Graph Neural Networks (Fi-GNN).
Taking advantage of the strong representative power of graphs, our proposed model can not only model sophisticated feature interactions in a flexible and explicit fashion, but also provide good model explanations for CTR prediction.
%The graph structure have intrinsic advantage of reflecting the interactions among different fields.
%We thus design a novel model Feature Interaction Graph Neural Networks (Fi-GNN) to model sophisticated interactions among feature fields in the form of node interactions on the corresponding graph, which is more effective and explicit. 
%Moreover, it can also provide good model explanations for CTR prediction.
Experimental results on two real-world datasets show its superiority over the state-of-the-arts.
\end{abstract}



%
% Keywords. The author(s) should pick words that accurately describe the work being
% presented. Separate the keywords with commas.


%
% This command processes the author and affiliation and title information and builds
% the first part of the formatted document.
\maketitle

\section{Introduction}
%Features play a central role in the success of many predictive systems.
%Distinct from continuous features which can be naturally found in images and audios, the features for web applications are mostly discrete and categorical.
%For example, in the task of Click-through rate (CTR) prediction, we need to predict how likely a user will click on an ad with four categorical variables: 1) user id = \{100, 101, 102, ...\}, 2) ad id = \{1001, 1002, 1003, ...\}, 3) advertisers = \{nike, nestle, apple, ...\} and 4) ad position = \{top, bottom, ...\}
%To build predictive models with these categorical features, a common solution is to convert them to a set of binary features via one-hot encoding. 
%Nevertheless, the generated feature vector are usually very high dimensional and sparse.
%A popular approach to dealing with such sparse data is Factorization Machines (FMs) \cite{Rendle2011Factorization}, which embed sparse features into low-dimensional dense vectors and learn feature interactions from vector inner products. 
%Field-aware factorization machines (FFMs) \cite{Juan2016Field} consider the field information and allow each feature to learn several vectors where each vector is associated with a field.
The goal of click-through rate prediction is to predict the probabilities of users clicking ads or items, which is critical to many web applications such as online advertising and recommender systems.
Modeling sophisticated feature interactions plays a central role in the success of CTR prediction.
Distinct from continuous features which can be naturally found in images and audios, the features for web applications are mostly in multi-field categorical form.
%For example, let us consider the task of Click-through rate (CTR) prediction, we need to predict how likely a user will click on an ad with four categorical variables: 1) user id = \{100, 101, 102, ...\}, 2) ad id = \{1001, 1002, 1003, ...\}, 3) advertisers = \{nike, nestle, apple, ...\} and 4) ad position = \{top, bottom, ...\} (noted that there are more features in real applications).
For example, the four-fields categorical features for movies may be: (1) \textsf{Language = \{English, Chinese, Japanese, ... \}}, (2) \textsf{Genre = \{action, fiction, ... \}}, (3) \textsf{Director = \{Ang Lee, Christopher Nolan, ... \}}, and (4) \textsf{Starring = \{Bruce Lee, Leonardo DiCaprio, ... \}} (noted that there are much more feature fields in real applications).
These multi-field categorical features are usually converted to sparse one-hot encoding vectors, and then embedded to dense real-value vectors, which can be used to model feature interactions.           
%When performing prediction on these categorical predictor variables, it is important to account for their interactions.
%There has been many efforts in modeling different orders of feature interactions.

Factorization machine (FM) \cite{rendle2010factorization} is a well-known model proposed to learn second-order feature interactions from vector inner products. 
Field-aware factorization machine (FFM) \cite{juan2016field} further considers the field information and introduces field-aware embedding.
Regrettably, these FM-based models can only model second-order interaction and the linearity modeling limits its representative power.      
%Although higher-order FMs have been proposed \cite{blondel2016higher}, they still belong to the family of linear models. 
Recently, many deep learning based models have been proposed to learn high-order feature interactions, which follow a general paradigm: simply concatenate the field embedding vectors together and feed them into DNN or other specifically designed models to learn interactions.
For example, Factorisation-machine supported Neural Networks (FNN) \cite{zhang2016deep}, Neural Factorization Machine (NFM) \cite{he2017neural}, Wide\&Deep \cite{cheng2016wide} and DeepFM \cite{guo2017deepfm} utilize DNN to model interactions.
However, these model based on DNN learn high-order feature interactions in a bit-wise, implicit fashion, which lacks good model explanations.
%In addition, the final function learned by DNNs can be arbitrary, and there is no theoretical conclusion on whether the interactions are effective enough. 
%In addition, they model feature interactions at the bit-wise level instead of vector-wise level in the traditional FM framework, which make it less interpretable. 
%Therefore, how effective are DNNs in modeling high-order feature interactions remains an open question. 
%xDeep \cite{lian2018xdeepfm} is thus proposed to solve the problem by combining the explicit and implicit feature interactions together.
%Some models such as Deep\&Cross \cite{wang2017deep} and xDeepFM \cite{lian2018xdeepfm} try to learn high order interactions explicitly by introducing specifically designed networks Cross Network (CrossNet) and Compressed Interaction Network (CIN) respectively.
Some models try to learn high order interactions explicitly by introducing specifically designed networks.
For example, Deep\&Cross \cite{wang2017deep} introduces Cross Network (CrossNet) and xDeepFM \cite{lian2018xdeepfm} introduces Compressed Interaction Network (CIN).
Nevertheless, we argue that they are still not sufficiently effective and explicit, since they still follow the general paradigm of combining feature fields together to model their interactions.
%In this way, the interaction function applied on different feature fields will inevitably be same.
The simple \emph{unstructured combination} will inevitably limit the capability to model sophisticated interactions among different feature fields in a flexible and explicit fashion.

In this work, we take the structure of multi-field features into consideration.
Specifically, we represent the multi-field features in a graph structure named \emph{feature graph}.
Intuitively, each node in the graph corresponds to a feature field and different fields can interact through edges.
%Since each two fields ought to interact in two directions, it's a complete weighted graph where the edge weights represent importances of different feature interactions.
%The graph structure has strong representative power since it can reveal the relations between different feature fields intrinsically.
%The graph-structured representations of categorical features has strong representative power.
The task of modeling sophisticated interactions among feature fields can be thus converted to modeling node interactions on the feature graph.
%GNN is designed to handle graph-structured data and has intrinsic advantages in modeling node interactions on the graphs.
To this end, we design a novel model Feature interaction Graph Neural Networks (Fi-GNN) based on Graph Neural Networks (GNN), which is able to model sophisticated node (feature) interactions in a flexible and explicit fashion.
%In Fi-GNN, the nodes will interact and update their state vectors in a recurrent fashion.
%At each interaction step, all the nodes interact with neighbors by communicating state information, and then update their states simultaneously.
%%The number of interaction steps equals to the hop of interacted neighbors, which is also the order of feature interactions.
%In Fi-GNN, the nodes will interact in a recurrent fashion.
%Specifically, the nodes will interact with their neighbors simultaneously at each interaction step.
In Fi-GNN, the nodes will interact by communicating the node states with neighbors and update themselves in a recurrent fashion.
At every time step, the model interact with neighbors at one hop deeper.  
Therefore, the number of interaction steps equals to the order of feature interactions.
Moreover, the edge weights reflecting importances of different feature interactions and node weights reflecting importances of each feature field on the final CTR prediction can be learnt by Fi-GNN, which can provide good explanations. 
%Moreover, the edge weights reflecting importances of different feature interactions can be learnt by Fi-GNN, which can provide good explanations for CTR prediction. 
Overall, our proposed model can model sophisticated feature interactions in an explicit, flexible fashion and also provide good model explanations.

%Otherwise, it can provide good model explanations, which are beneficial to help us understand CTR prediction, especially recommender systems.

%Fi-GNN will propagate the state information 
%At each propagation step, the nodes interacts with their neighbors simultaneously and then update the state vectors according the aggregated neighbors' states, via GRU and residual connection.
%That is to say, our proposition can model certain order feature interactions, which equals to the propagation step, in an explicit, vector-wise fashion.

%can model certain order feature interactions in a more explicit and efficient fashion compared with conventional methods.

Our contributions can be summarized in threefold:
\begin{itemize}
\item 
We point out the limitation of the existing works which consider multi-field features as an unstructured combination of feature fields.
To this end, we propose to represent the multi-field features in a graph structure for the first time.  
%\item We design a GNN-based model Fi-GNN to model certain order feature interactions on the graph-structured features in an explicit, vector-wise fashion.
\item We design a novel model Feature Interaction Graph Neural Networks (Fi-GNN) to model sophisticated interactions among feature fields on the graph-structured features in a more flexible and explicit fashion.
\item Extensive experiments on two real-world datasets show that our proposed method can not only outperform the state-of-the-arts but also provide good model explanations. 
\end{itemize}

The rest of this paper is organized as follows. 
%Section 2 provides some preliminary knowledge to help understand our proposed model and Section 3 introduces it in detail. 
Section 2 summarizes the related work.
Section 3 provides an elaborative description of our proposed method.
The extensive experiments and detailed analysis are presented in Section 4, followed by the conclusion in Section 5.

\section{Related Work}
In this section, we briefly review the existing models that model feature interactions for CTR prediction and graph neural networks.
\subsection{Feature Interaction in CTR Prediction} \label{sect:related}
Modeling feature interactions is the key to success of CTR prediction and therefore extensively studied in the literature.
%The structure of CTR prediction model has evolved from shallow to deep.
%There are mainly three types of existing base models to model feature interactions: 
%(1) Logistic Regression (LR) which models first-order interaction;
%(2) Factorization Machine (FM) based linear models which model second-order interactions; 
%(3) Deep Neural Networks (DNN) based non-linear models which model high-order interactions.
LR is a linear approach, which can only models the first-order interaction on the linear combination of raw individual features.
FM \cite{rendle2010factorization} learns second-order feature interactions from vector inner products. 
Afterwards, different variants of FM have been proposed.
Field-aware factorization machine (FFM) \cite{juan2016field} considers the field information and introduces field-aware embedding.
AFM \cite{xiao2017attentional} considers the weight of different second-order feature interactions.
However, these approaches can only model second-order interaction which is not sufficient.
%, since the real-world data usually have complex and non-linear underlying relations.
%Although higher-order FMs \cite{blondel2016higher} have been proposed, they still belong to the family of linear models. 

%To learn high-order feature interactions, researchers have proposed many approaches, which follow a general paradigm:  concatenate the field embedding vectors together and feed them into DNN or other specifically designed networks to learn the nonlinear high-order interaction.
%Factorisation-machine supported Neural Networks (FNNs) \cite{zhang2016deep}, Product-based Neural Network (PNN) \cite{qu2016product}, Neural Factorization Machine (NFM) \cite{he2017neural}, Wide\&Deep \cite{cheng2016wide} and DeepFM \cite{guo2017deepfm} are the representative works which utilize DNN to model high-order interactions.
%Regrettably, all these approaches leveraging DNN learn the high-order feature interactions in an implicit, bit-wise way and therefore lack good model explainability. 
%Recently, some works try to learn feature interactions in an explicit fashion via specifically designed networks. 
%Deep\&Cross \cite{wang2017deep} introduces a CrossNet which takes outer product of features at the bit level.
%On the contrary, xDeepFM \cite{lian2018xdeepfm} introduces a CIN to take outer product at the vector level.
With the success of DNN in various fields, researchers start to use it to learn high-order feature interactions due to its deeper structures and nonlinear activation functions.
The general paradigm is to concatenate the field embedding vectors together and feed them into DNN to learn the high-order feature interactions.
\cite{liu2015convolutional} utilizes convolutional networks to model feature interactions. 
Factorisation-machine supported Neural Networks (FNNs) \cite{zhang2016deep} uses the pre-trained factorization machines for field embedding before applying DNN.
Product-based Neural Network (PNN) \cite{qu2016product} models both second-order and high-order interactions by introducing a product layer between field embedding layer and DNN layer.
%, and does not rely on pre-trained FM.
Similarly, Neural Factorization Machine (NFM) \cite{he2017neural} has a Bi-Interaction Pooling layer between embedding layer and DNN layer to model second-order interactions, but the followed operation is summation instead of concatenation as in PNN. 
%PNN and NFM have the problem of focusing more on high-order interactions but less on low-order, due to their series connection of the two base models.
Some works on another line try to model the second-order and high-order interactions jointly via a hybrid architectures.
The Wide\&Deep \cite{cheng2016wide} and DeepFM \cite{guo2017deepfm} contain a wide part to model the low-order interaction and a deep part to model the high-order interaction.
However, all these approaches leveraging DNN learn the high-order feature interactions in an implicit, bit-wise way and therefore lack good model explainability. 
Recently, some work try to learn feature interactions in an explicit fashion via specifically designed networks. 
Deep\&Cross \cite{wang2017deep} introduces a CrossNet which takes outer product of features at the bit level.
On the contrary, xDeepFM \cite{lian2018xdeepfm} introduces a CIN to take outer product at the vector level.
Nevertheless, they still don't solve the most fundamental problem, that is to concatenate the field embedding vectors together.
The simple unstructured combination of feature fields will inevitably limit the capability to model sophisticated interactions among different fields in a flexible and explicit fashion.
To this end, we proposed to represent the multi-field features in a graph structure, where each node represents a field and different feature fields can interact through the edges.
Accordingly, we can model the flexible interactions among different feature fields on the graphs. 
 
%The graph-structured representation can reveal the relations between features intrinsically.
%We then proposed a GNN-based model Fi-GNN to model the sophisticated features interaction on the graph-structured features.  
%Experimental results prove the effectiveness of graph-structured feature representation.

\subsection{Graph Neural Networks}
%We first provides some preliminary knowledge of GNNs before we introduce our proposed model Fi-GNN.
Graph is a kind of data structure which models a set of objects (nodes) and their relationships (edges). 
Recently, researches of analyzing graphs with machine learning have been receiving more and more attention because of the great representative power of graphs.
Early works usually convert graph-structured data into sequence-structured data to deal with. 
Inspired by word2vec \cite{mikolov2013distributed},
the work \cite{perozzi2014deepwalk} proposed an unsupervised DeepWalk algorithm to learn node embedding in graph based on random walks. 
After that, \cite{tang2015line} proposed a network embedding algorithm LINE, which preserve the first- and second-order structural information.
\cite{Grover2016node2vec} proposed node2vec which introduces a biased random walk.
However, these methods can be computationally expensive and non-optimal for large graphs. 

Graph neural networks (GNN) are designed to tackle these problems, which are deep learning based methods that operate on the graph domain.
The concept of GNN is first proposed by \cite{scarselli2009graph}.
%There have been various kinds of GNN proposed these days.
Generally, nodes in GNNs interact with neighbors by aggregating information from neighborhoods and updating their hidden states.
There have been many variants of GNN with various kinds of aggregators and updaters proposed these days.
%Some surveys \cite{wu2019comprehensive,zhou2018graph} provide thorough reviews on them. 
Here we only present some representative and classical methods.
Gated Graph Neural Networks (GGNN) \cite{li2015gated} uses GRU \cite{cho2014learning} as updater.
Graph Convolutional Networks (GCN) \cite{kipf2016semi} considers the spectral structure of graphs and utilizes the convolutional aggregator.
GraphSAGE \cite{hamilton2017inductive} considers the spatial information. It introduces three kinds of aggregators:
mean aggregator, LSTM aggregator and Pooling aggregator.
Graph attention network (GAT) \cite{velivckovic2017graph} incorporates the attention mechanism into the propagation step.
There are some surveys \cite{wu2019comprehensive,zhou2018graph} which provide more elaborative introduction of various kinds of GNN models.

\begin{figure}[t]
\centering
\includegraphics[width=0.95\linewidth]{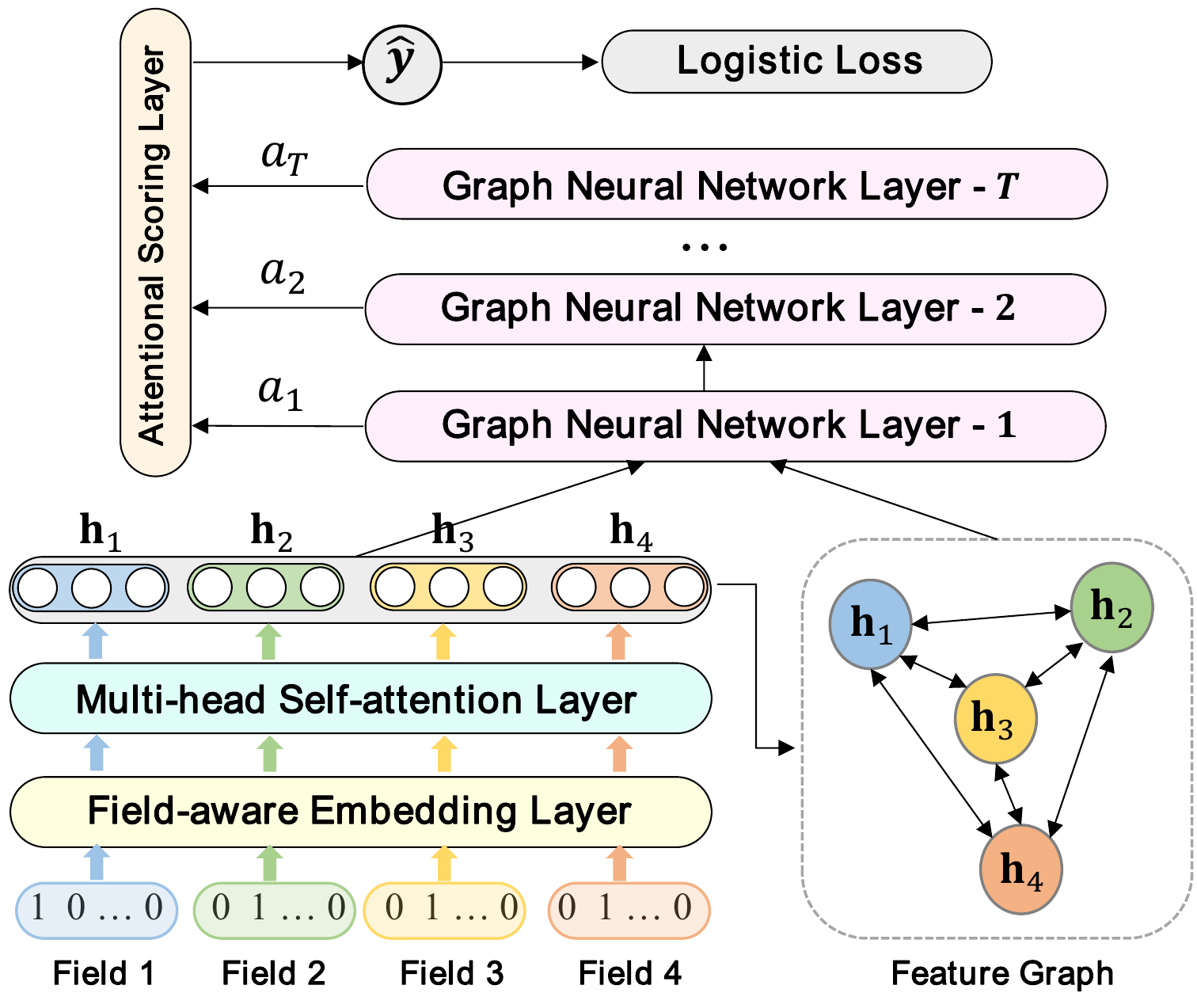}
%\vspace{-2mm}
\caption{
Overview of our proposed method.
The input raw multi-field feature vector is first converted to field embedding vectors via an embedding layer and represented as a feature graph, which is then feed into Fi-GNN to model feature interactions.
An attention layer is applied on the output of Fi-GNN to predict the click through rate $\hat{y}$.
Details of embedding layer and Fi-GNN are illustrated in Figure 2 and Figure 3 respectively.}
\vspace{-4mm}
\label{fig:overview}
\end{figure}

Due to convincing performance and high interpretability, GNN has been a widely applied graph analysis method.
Recently, there are many application of GNN like neural machine translation \cite{beck2018graph}, semantic segmentation \cite{qi20173d}, image classification \cite{marino2017more}, situation recognition \cite{li2017situation}, recommendation \cite{Wu2018Session}, script event prediction \cite{Zhongyang2018Constructing}, fashion analysis \cite{cui2019dressing,li2019semi}.
GNN is suitable for modeling node interactions on graph-structured features intrinsically.
In this work, we proposed a model Fi-GNN based on GGNN to model feature interactions on the graph-structured features for CTR prediction. 

\section{Our Proposed Method}
We first formulate the problem and then introduce the overview of our proposed method, followed by the elaborate detail of each component.
 
\subsection{Problem Formulation}
%Suppose the training dataset consists of instances $(\mathbf{x}, y)$, where $\mathbf{x}$ is a $m$-fields categorical feature ($m$ is the number of feature fields) and $y \in \left \{ 0,1 \right \}$ is the associated label indicating user click behaviors.
%The task of CTR prediction is to build a prediction model $\hat{y}$ = \emph{model}($\mathbf{x}$) to estimate the probability of a user clicking, whose key is to model the sophisticated interactions among different feature fields.
Suppose the training dataset consists of $m$-fields categorical features ($m$ is the number of feature fields) and the associated labels $y \in \left \{ 0,1 \right \}$ which indicate user click behaviors.
The task of CTR prediction is to predict $\hat{y}$ for the input $m$-fields features, which estimates the probability of a user clicking.
The key of the task is to model the sophisticated interactions among different feature fields.

\subsection{Overview}
Figure \ref{fig:overview} is the overview of our proposed method ($m$=4).   
The input sparse $m$-field feature vector is first mapped into sparse one-hot embedding vectors and then embedded to dense field embedding vectors via the embedding layer and the multi-head self-attention layer. 
The field embedding vectors are then represented as a feature graph, where each node corresponds to a feature field and different feature fields can interact through edges.
%Since each two feature fields ought to interact, it is a weighted complete graph.
The task of modeling interaction can be thus converted to modeling node interactions on the feature graph.
%\footnote{Node interactions equal to feature interactions in the following.}.
%, with edge weights (adjacency matrix) learnt automatically in Fi-GNN. 
Therefore, the feature graph is feed into our proposed Fi-GNN to model node interactions.  
An attention scoring layer is applied on the output of Fi-GNN to estimate the click-through rate $\hat{y}$. 
In the following, we will introduce the details of our proposed method.
    
%\begin{figure}[t]
%\centering
%\vspace{-2mm}
%%\setlength{\abovecaptionskip}{0pt}
%%\setlength{\belowcaptionskip}{-10pt}
%\includegraphics[width=1\linewidth]{./pic/embedding6.pdf}
%\caption{An example of the embedding layer.
%The raw feature is first mapped into sparse one-hot embedding vectors and then embedded to dense field embedding vectors via a field-aware embedding layer.}
%
%\label{fig:embedding}
%\end{figure}

\subsection{Embedding Layer} \label{sect:graph}
%In the field of computer vision or natural language processing,  the input features are usually continuous, which can be naturally found in images, audios or textual signal.
The multi-field categorical feature  $\mathbf{x}$ is usually sparse and of huge dimension.
Following previous works \cite{zhang2016deep,qu2016product,wang2017deep,guo2017deepfm,qu2018product}, we represent each field as a one-hot encoding vector and then embed it to a dense vector, noted as field embedding vector.
Let us consider the example in Section 1, 
%a English fiction movie directed by Christopher Nolan, starred by Leonardo DiCaprio 
a movie \textsf{\{Language: English, Genre: fiction, Director: Christopher Nolan, Starring: Leonardo DiCaprio \}} is first transformed into a high-dimensional sparse features via one-hot encoding:
\begin{center}	
$ \underbrace{\left [ 1, 0, ..., 0  \right ]}_{\text{Language}}, \underbrace{\left [ 0, 1, ..., 0 \right ]}_{\text{Genre}}, \underbrace{\left [ 0, 1, ..., 0  \right ]}_{\text{Director}}, \underbrace{\left [ 0, 1, ..., 0  \right ]}_{\text{Starring}}$
 \end{center}	
A field-aware embedding layer is then applied upon the one-hot vectors to embed them to low dimensional, dense real-value field embedding vectors as shown in Figure \ref{fig:embedding}.
Likewise, the field embedding vectors of $m$-field feature can be obtained: 
\begin{center}	
$ \mathbf{E} = \left [ \mathbf{e}_{1}, \mathbf{e}_{2}, \mathbf{e}_{3}, ..., \mathbf{e}_{m} \right ], $
\end{center}
where $\mathbf{e}_{i} \in \mathbb{R}^{d}$ denotes the embedding vector of field $i$ and $d$ denotes the dimension of field embedding vectors.

\subsection{Multi-head Self-attention Layer} 
Transformer~\cite{vaswani2017attention} is prevalent in NLP and has achieved great success in many tasks.
At the core of Transformer, the multi-head self-attention mechanism is able to model complicated dependencies between word pairs in multiple semantic subspaces.
In the literature of CTR prediction, we take advantage of the multi-head self-attention mechanism to capture the complex dependencies between feature field pairs, i.e, pairwise feature interactions, in different semantic subspaces.

Following~\cite{song2018autoint}, given the feature embeddings $\mathbf{E}$, we obtain the feature representation of features that cover the pairwise interactions of an attention head $i$ via scaled dot-product:
\begin{equation}\nonumber
\mathbf{H}_{i} = \text{softmax}_{i}(\frac{\mathbf{Q}\mathbf{K}^{T}}{\sqrt{d_{K}}})\mathbf{V},
\end{equation}
\begin{equation}\nonumber
\mathbf{Q}=\mathbf{W}_i^{(Q)}\mathbf{E}, \mathbf{K}=\mathbf{W}_i^{(K)}\mathbf{E}, \mathbf{V}=\mathbf{W}_i^{(V)}\mathbf{E}. 
\end{equation}
The matrices $\mathbf{W}_i^{(Q)} \in \mathbb{R}^{d_i \times d}$, $\mathbf{W}_i^{(K)} \in \mathbb{R}^{d_i \times d}$, $\mathbf{W}_i^{(V)} \in \mathbb{R}^{d_i \times d}$ are three weight parameters for attention head $i$, $d_i$ is the dimension size of head $i$, and $\mathbf{H}_{i} \in \mathbb{R}^{m \times d_i}$.

Then we combine the learnt feature representations of each head to preserve the pairwise feature interactions in each semantic subspace:
\begin{equation}\nonumber
\mathbf{H}^1 = \text{ReLU}(\mathbf{H}_{1}\oplus \mathbf{H}_{2}\oplus \cdots \oplus \mathbf{H}_{h}),
\end{equation}
where $\oplus$ denotes the concatenation operation and $h$ denotes the number of attention heads.
The learnt feature representations $\mathbf{H}^1 \in \mathbb{R}^{m \times d'}$ are used for the initial node states of the graph neural network, where $d' = \sum_{i=1}^h d_i$.

%\subsection{Graph-structured Feature Representation} 
\subsection{Feature Graph} 
%Distinguished from the previous works focusing on improving structures of models, we focus more on the structure of multi-field feature, which is more fundamental.
Distinguished from the previous works which simply concatenate the field embedding vectors together and feed them into designed models to learn feature interactions, we represent them in a graph structure.
In particular, We represent each input multi-field feature as a \emph{feature graph} $\mathcal{G} = (\mathcal{N}, \mathcal{E})$,
where each node $n_{i} \in \mathcal{N}$ corresponds to a feature field $i$ and different fields can interact through the edges, so that $\left | \mathcal{N} \right | = m$. 
Since each two fields ought to interact, it is a weighted fully connected graph while the edge weights reflect importances of different feature interactions. 
%The graph structure has a strong representative power, which can reveal the relations between different feature fields intrinsically.
Accordingly, the task of modeling feature interactions can be converted to modeling node interactions on the feature graph. 
%To this end, we proposed Fi-GNN as introduced in Section \ref{sect:model}.
%We then use our proposed Fi-GNN to model feature interactions on the graph-structured features, as introduced in Section \ref{sect:model}.

%The field embedding $e_{i}$ is utilized as the the initial hidden state of node $n_{i}$ in GNNs.
%Since features of any two fields should interact with each other, the graphs are fully connected. 
%The weights of edges (i.e. adjacency matrix) are learnt in our model, which can also reflect the relations between features.
%In GNNs, each node can interact with all its neighbor simultaneously.
%Specularly, the state vector of each node is updated by aggregating all its neighbors states. 
%That is to say, our proposition can model the certain bound-degree interaction in an explicit, vector-wise fashion.

\subsection{Feature Interaction Graph Neural Network}\label{sect:model} 
Fi-GNN is designed to model node interactions on the feature graph, which is based on GGNN \cite{li2015gated}.
It is able to model the interactions in a flexible and explicit fashion.

\begin{figure}[t]
\centering
\includegraphics[width=1\linewidth]{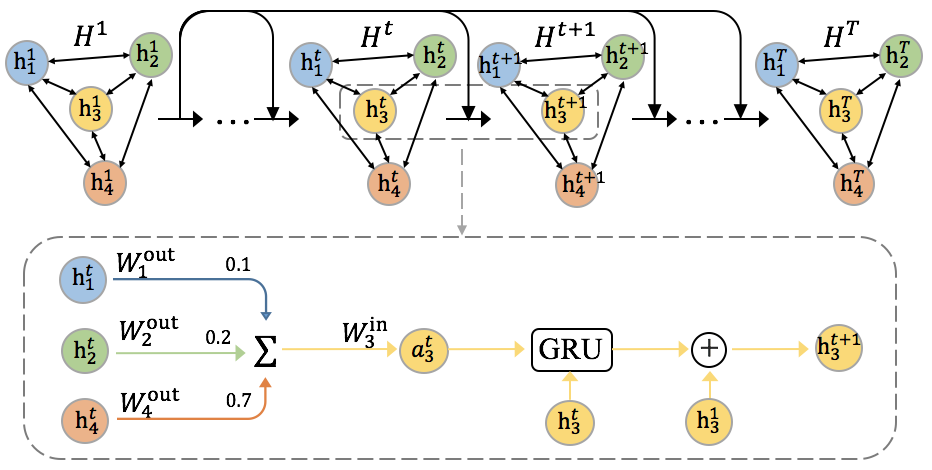}
\caption{Framework of Fi-GNN. 
The nodes interact with neighbors and update their states in a recurrent fashion.
At each interaction step, each node will first aggregate transformed state information from neighbors and then update its state according to the aggregated information and history via GRU and residual connection.}
%The input could be image, text or any other modality of items.}
\label{fig:framework}
\end{figure}

\noindent \textbf{Preliminaries.}
In Fi-GNN, each node $n_{i}$ is associated with a hidden state vector $\mathbf{h}_{i}^{t}$ and the state of graph is composed of these node states
\begin{center}	
$\mathbf{H}^{t} = \left [ \mathbf{h}_{1}^{t}, \mathbf{h}_{2}^{t}, \mathbf{h}_{3}^{t}, ... ,
\mathbf {h}_{m}^{t} \right ],$
\end{center}
where $t$ denote the interaction step.
The learnt feature representations by the multi-head self-attention layer are used for their initial node states $\mathbf{H}^{1}$.
As shown in Figure \ref{fig:framework}, the nodes interact and update their states in a recurrent fashion. 
At each interaction step, the nodes aggregate the transformed state information with neighbors, and then update their node states according to the aggregated information and history via GRU and residual connection.
Next, we will introduce the details of Fi-GNN elaborately.

%\noindent \textbf{Initial Node States.}
%The initial node states are that of the input feature graphs, i.e., the field embedding vectors.
%In particular, the embedding vector $e_{i}$ of field $i$ is utilized as the initial state vector of the corresponding node $n_{i}$, which can be formalized as,
%\begin{equation}
%\mathbf{H}^{1} = \mathbf{E},
%\end{equation}
%\begin{equation}
%\mathbf{h}_{i}^{1} = \mathbf{e}_{i},   i \in \left[1,2, ..., m\right].
%\end{equation}
%Accordingly, $\mathbf{h} \in \mathbb{R}^{D}, \mathbf{H} \in \mathbb{R}^{m \times D}.$

\noindent \textbf{State Aggregation.}
At interaction step $t$, each node will aggregate the state information from neighbors.
Formally, the aggregated information of node $n_{i}$ is sum of its neighbors' transformed state information,
\begin{equation} \label{ggnn_original}
%\mathbf{a}_{i}^{t} = \sum_{j} \mathbf{A}^{T}h^{t-1} + b,
\mathbf{a}_{i}^{t} = \sum_{n_{j} \rightarrow n_{i} \in \mathcal{E}} \mathbf{A}[n_{j}, n_{i}] \mathbf{W}_{p} \mathbf{h}_{j}^{t-1},~
\end{equation}
where $\mathbf{W}_{p}$ is the transformation function.
$\mathbf{A} \in \mathbb{R}^{m \times m}$ is the adjacency matrix containing the edge weights.
For example, $\mathbf{A}[n_{j}, n_{i}]$ is the weight of edge from node $n_{j}$ to $n_{i}$, which can reflect the importance of their interaction.
Apparently, the transformation function and adjacency matrix decide on the node interactions.
Since the interaction on each edge ought to differ, we aim to achieve edge-wise interaction, which requires a unique weight and transformation function for each edge.        
%There exist two limitations in the conventional GGNN, which make it fail to model flexible node interactions.
%(1) \textit{\textbf{static edge weights}}:
%since the edge weights of graphs are expensive to obtain, the unique adjacency matrix for each graph is the adjacency matrix is usually given in the binary form, or a static one pre-calculated from the statistical data;
%(2) \textit{\textbf{shared state transformation}}: the state information is transformed in a fixed way via a shared state transformation function on all the edges.

(1) \textit{\textbf{Attentional Edge Weights.}}
%\noindent \textbf{Attentional Edge weights.}
The adjacency matrix in the conventional GNN models is usually in the binary form, i.e., only contains 0 and 1.
It can only reflect the connected relation of nodes but fails to reflect the importances of their relations.
In order to infer the importances of interactions between different nodes, we propose to learn the edge weights via an attention mechanism.
%Similar with GAT \cite{velivckovic2017graph}, the weight of edge from node $n_{i}$ to node $n_{j}$ is calculated with their node states.
In particular, the weight of edge from node $n_{i}$ to node $n_{j}$ is calculated with their initial node states, i.e., the corresponding field embedding vectors.
Formally,
%\begin{equation} \label{uniform_A}
%w(n_{i}, n_{j}) =  \sigma(\mathbf{W}_{w} \left [ \mathbf{e}_{i} \left |  \right | \mathbf{e}_{j} \right ] + b_{w}),
%\end{equation}
%\begin{equation} \label{uniform_A}
%c(n_{i}, n_{j}) =  \text{LeakyRelu}(\mathbf{W}_{w} \left [ \mathbf{e}_{i} \left |  \right | \mathbf{e}_{j} \right ]),
%\end{equation}
\begin{equation} \label{uniform_A}
w(n_{i}, n_{j}) = \frac{\text{exp}(\text{LeakyRelu}(\mathbf{W}_{w} \left [ \mathbf{e}_{i} \left |  \right | \mathbf{e}_{j} \right ])) }{\sum_{k} \text{exp}(\text{LeakyRelu}(\mathbf{W}_{w} \left [ \mathbf{e}_{i} \left |  \right | \mathbf{e}_{k} \right ]))},
\end{equation}
where $\mathbf{W}_{w} \in \mathbb{R}^{2d'}$ is a weight matrix, $\left |  \right |$ is the concatenation operation.
The softmax function is utilized to make weights easily comparable across different nodes.
Therefore, the adjacency matrix is,
\begin{equation} \label{a}
\mathbf{A}[n_{i}, n_{j}]=\begin{cases}
 & w(n_{i}, n_{j}),  \text{ if } i \neq j, \\
 & 0, \text{ else }.
\end{cases}
\end{equation}
%Noted that the way we calculate the edge weight is similar to GAT \cite{velivckovic2017graph}, which we experimentally found achieve worse
%performance than ours.

Since the edge weights reflects the importances of different interaction, Fi-GNN can provide good explanations on the relation of different feature fields of input instance, which will be further discussed in Section \ref{sect:explannation}.
%Since the weight of each edge between two nodes (fields) reflects the importance of their interaction, Fi-GNN can provide good explanations on the relation of different feature fields of input instance, which will be further discussed in Section \ref{sect:explannation}.

%$\mathbf{A} \in \mathbb{R}^{m \times m}$ is the adjacency matrix and
%$\mathbf{A}[n_{j}, n_{i}]$ is the weight of edge between node $n_{j}$ and $n_{i}$, which reflects the strengthen of their relation.
%Higher the weight is, stronger their relation is.
%In the traditional GGNNs model, the adjacency matrix is given.
%However, the relation between nodes is usually unknown in real world applications.
%Here we propose to learn the weights automatically as,
%\begin{equation} \label{uniform_A}
%\mathbf{A} = HW_{w},
%\end{equation}
%where $H \in \mathbb{R}^{m \times d}$ is the state of nodes and $w_{w} \in \mathbb{R}^{d \times m}$ is a weight matrix.
%The traditional GGNN use shared transformation weight $\mathbf{W}_{p}$ and bias $\mathbf{b}_{p}$ on all edges to model interactions between different nodes, which fails to model complex edge-wise interactions.

(2) \textit{\textbf{Edge-wise Transformation.}}
%The nodes interact by communicating their transformed state information through the edge, so that the transformation function decide the interaction fashion.
%Apparently, a fixed transformed function on all the edges is unable to model the flexible interactions.
%Since the interaction between each two nodes, i.e., the transformation of state information on each edge, is supposed to be different, the edge-wise transformation function is essential.  
As discussed before, a fixed transformed function on all the edges is unable to model the flexible interactions and a unique transformation for each edge is essential. 
Nevertheless, our graph is complete graph with a huge number of edges.
Simply assigning a unique transformation weight to each edge will consuming too much parameter space and running time. 
To reduce the time and space complexity and also achieve edge-wise transformation, we assign an output matrix $\mathbf{W}_{out}^{i}$ and an input matrix $\mathbf{W}_{in}^{i}$ to each node $n_{i}$ similar with \cite{cui2019dressing}.
As shown in Figure \ref{fig:framework}, when node $n_{i}$ sends its state information to node $n_{j}$, the state information will first be transformed by its output matrix $\mathbf{W}_{out}^{i}$ and then transformed by node $n_{j}$'s input matrix $\mathbf{W}_{in}^{j}$ before $n_{j}$ receives it. 
The transformation function of edge $n_{i} \rightarrow n_{j}$ from node $n_{i}$ to node $n_{j}$ thus could be written as,
\begin{equation} \label{W_inout}
\mathbf{W}_{p}^{n_{i} \rightarrow n_{j}} = \mathbf{W}_{out}^{i}\mathbf{W}_{in}^{j}.
\end{equation}
Likewise, the transformation function of edge $n_{j} \rightarrow n_{i}$ from node $n_{j}$ to node $n_{j}$ is
\begin{equation} \label{W_inout}
\mathbf{W}_{p}^{n_{j} \rightarrow n_{i}} = \mathbf{W}_{out}^{j}\mathbf{W}_{in}^{i}.
\end{equation}
%Since the node interaction on each edge ought to be different, we aim to model edge-wise interaction. There is a straight-forward way to achieve edge-wise node interactions, that is to let each edge has its own transformation weight and bias $W_{p}$ and $b_{p}$. However, it will cost a great parameter space with numerous edges in a graph and cannot be applied on large scale graphs.
Accordingly, the Equation \ref{ggnn_original} could be rewritten as, % at each propagation step $t$, the node $n_{i}$ will receive information $\mathbf{a}_{i}^{t}$ from its neighbors as,
\begin{equation} \label{ggnn_1}
\mathbf{a}_{i}^{t} = \sum_{n_{j} \rightarrow n_{i} \in \mathcal{E}} \textbf{A}[n_{j}, n_{i}]\mathbf{W}_{out}^{j}\mathbf{W}_{in}^{i} \mathbf{h}_{j}^{t-1} + \mathbf{b}_{p}.
\end{equation}
In this way, the number of parameters is proportional to the number of nodes rather than numerous edges, which greatly reduces the space and time complexity and meanwhile achieves edge-wise interaction.

\noindent \textbf{State Update.}
After aggregating state information, the nodes will update the state vectors via GRU and residual connections.

(1) \textit{\textbf{State update via GRU.}}
In traditional GGNN, the state vector of node $n_{i}$ is updated via GRU based on the aggregated state information $\mathbf{a}_{i}^{t}$ and its state at last step.
Formally, 
\begin{equation} \label{ggnn_1}
\mathbf{h}_{i}^{t} = GRU(\mathbf{h}_{i}^{t-1}, \mathbf{a}_{i}^{t}).
\end{equation}
It can be formalized in detail as:
%\begin{align}
%& \mathbf{z}_{i}^{t} =  \sigma(W_{z} a_{i}^{t} + U_{z}h_{i}^{t-1} + b_{z}),\\             
%& \mathbf{r}_{i}^{t} =  \sigma(W_{r} a_{i}^{t} + U_{r}h_{i}^{t-1} + b_{r}),\\
%& \mathbf{\tilde{h}}_{i}^{t} =  tanh(W_{h} a_{i}^{t} + U_{h}(\mathbf{r}_{i}^{t} \odot h_{i}^{t-1}) + b_{h}), \\
%& \mathbf{h}_{i}^{t} = \mathbf{\tilde{h}}_{i}^{t} \odot z_{i}^{t} + \mathbf{h}_{i}^{t-1} \odot (1-z_{i}^{t}).
%\end{align}
\begin{align}
& \mathbf{z}_{i}^{t} =  \sigma(\mathbf{W}_{z} \mathbf{a}_{i}^{t} + \mathbf{U}_{z}\mathbf{h}_{i}^{t-1} + \mathbf{b}_{z}),\\
& \mathbf{r}_{i}^{t} =  \sigma(\mathbf{W}_{r} \mathbf{a}_{i}^{t} + \mathbf{U}_{r}\mathbf{h}_{i}^{t-1} + \mathbf{b}_{r}),\\
& \mathbf{\tilde{h}}_{i}^{t} =  tanh(\mathbf{W}_{h} \mathbf{a}_{i}^{t} + \mathbf{U}_{h}(\mathbf{r}_{i}^{t} \odot \mathbf{h}_{i}^{t-1}) + \mathbf{b}_{h}), \\
& \mathbf{h}_{i}^{t} = \mathbf{\tilde{h}}_{i}^{t} \odot \mathbf{z}_{i}^{t} + \mathbf{h}_{i}^{t-1} \odot (1-\mathbf{z}_{i}^{t}),~
\end{align}
where, $\mathbf{W}_{z}$, $\mathbf{W}_{r}$, $\mathbf{W}_{h}$, $\mathbf{b}_{z}$, $\mathbf{b}_{r}$, $\mathbf{b}_{h}$ are weights and biases of the updating function Gated Recurrent Unit (GRU) \cite{li2015gated}. 
$\mathbf{z}_{i}^{t}$ and $\mathbf{r}_{i}^{t}$ are update gate vector and reset gate vector, respectively.
%After each propagation step, we zero the hidden state of the unconcerned nodes making them not receiving or sending information. 
 
(2) \textit{\textbf{State update via Residual Connections.}}
Previous works \cite{shan2016deep,song2018autoint,cheng2016wide} have proved that it's effective to combine the low-order and high-order interactions together.
We thus introduce extra residual connections to update note states along with GRU, which can facilitate low-order feature reuse and gradients back-propagation.
Therefore, the Eq. (\ref{ggnn_1}) can be rewritten as, 
\begin{equation} \label{eq:residual}
\mathbf{h}_{i}^{t} = GRU(\mathbf{h}_{i}^{t-1}, \mathbf{a}_{i}^{t}) + \mathbf{h}_{i}^{1}.
\end{equation}
%Noted that the original residual connection first proposed in the Residual Net \cite{he2016deep} is formalized as $y=F(x)+x$, we make some adjustment here, which is experimentally proven to be more effective.
%\begin{equation} 
%\mathbf{h}_{i}^{t} = GRU(\mathbf{h}_{i}^{t-1}, \mathbf{a}_{i}^{t}) + \mathbf{h}_{i}^{t-1}.
%\end{equation}
%However, we experimentally found that it achieves worse performance compared with Equation \ref{eq:residual}.
%As far as I'm concerned, this may due to that $\mathbf{h}_{i}^{t-1}$ is redundant while $\mathbf{h}_{i}^{1}$ is sufficient to provide low-order information.

\subsection{Attentional Scoring Layer}
After $T$ propagation steps, we can obtain the node states
\begin{center}
$\mathbf{H}^{T} = \left [ \mathbf{h}_{1}^{T}, \mathbf{h}_{2}^{T}, ... , \mathbf{h}_{m}^{T} \right ].$
\end{center}
Since the nodes have interacted with their $T$-order neighbors, the $T$-order feature interactions is modeled.
%Before we make predictions on these final node states, we should first know what they exactly are.
We need a graph-level output to predict CTR.

\noindent{\textbf{Attentional Node Weights}}
The final state of each field node has captured the global information. In other words, these field nodes are neighborhood-aware. 
%which can be thus used for prediction directly.
Here we predict a score on the final state of each field respectively and sum them up with an attention mechanism which measures their influences on the overall prediction.
Formally, the prediction score of each node $n_{i}$ and its attentional node weight can be estimated via two multiple layers perceptions respectively as,
\begin{equation} \label{self_attention1}
%\hat{y}_{i} = \sigma(\theta(\mathbf{h}_{i}^{p})),
\hat{y}_{i} = MLP_{1}(\mathbf{h}_{i}^{p})),
\end{equation}
\begin{equation} \label{self_attention2}
%\hat{y}_{i} = \sigma(\theta(\mathbf{h}_{i}^{p})),
a_{i} = MLP_{2}(\mathbf{h}_{i}^{p})).
\end{equation}
The overall prediction is a summation of all nodes:
%\begin{equation} \label{self_attention}
%\hat{y} = \sum_{i=1}^{m} \sigma(\theta(\mathbf{h}_{i}^{p})) \cdot \alpha(\delta(\mathbf{h}_{i}^{p})).
%\end{equation}
\begin{equation} \label{self_attention}
\hat{y} = \sum_{i=1}^{m}a_{i}\hat{y}_{i}.
\end{equation}
Note that it is actually same as the work \cite{li2015gated}.
%$\theta(\cdot)$ and $\delta(\cdot)$ are two perception networks to output a real value.
%$\theta(\cdot)$ is used to model the weights of each field (i.e., importance of fields' influence on the overall prediction score), and $\delta(\cdot)$ is used to model the prediction score of each field respectively.
%$\alpha (\cdot)$ are leaky relu activate functions $\alpha(x)=max(0.01x, x)$ and $\sigma(\cdot)$ is sigmoid activate functions $\sigma(x)=1/(1+e^{-x})$.
Intuitively, $MLP_{1}$ is used to model the prediction score of each field aware of the global information and $MLP_{2}$ is used to model the weights of each field (i.e., importance of fields' influence on the overall prediction).

%To make it intuitive, let us consider the example of movies.
%The final state of field \textsf{Starring} is already aware of other fields \textsf{Language}, \textsf{Director} and \textsf{Genre}, so are other fields.
%Therefore, we can predict scores on these fields and calculate their importance on the overall prediction score respectively, and finally sum them up.

\subsection{Training}
Our loss function is Log loss, which is defined as follows:
\begin{equation}
\mathcal{L} = -\frac{1}{N} \sum_{i=1}^{N}(y_{i}log(\hat{y}_{i})+(1-y_{i})log(1-\hat{y}_{i})),
\end{equation} \label{eqa:logloss}
where $N$ is the total number of training samples and $i$ indexes the training samples.
The parameters are updated via minimizing the Log Loss using RMSProp \cite{tieleman2012lecture}.
Most CTR datasets have unbalanced proportion of positive and negative samples, which will mislead the predictions.
To balance the proportion, we randomly select
 equal number of positive and negative samples in each batch during training process.
 
\subsubsection{\textbf{Parameter Space.}}
The parameter needed to be learnt mainly consists of the parameters correlated to nodes and the perception networks in attention mechanism.
%The main difference between NGNN and GGNN is in the node interaction parameter.
For each node $n_{i}$, we have an input matrix $\mathbf{W}_{in}^{i}$ and an output matrix $\mathbf{W}_{out}^{i}$ to transform state information. 
Totally we have $2m$ matrices, which are proportional to the number of nodes $m$. 
Besides, the multi-head self-attention layer contains the following weight matrices $\left \{    \mathbf{W}_i^{(Q)},   \mathbf{W}_i^{(K)}, \mathbf{W}_i^{(V)}  \right \}$ for each head, and the number of parameters of the entire layer is $(3dd'+hdd')$.  
In addition, we have two matrices of perception networks in the self-attention mechanism and also parameters in GRU. 
Overall, there are $O(2m+hdd')$ matrices. 

%\subsubsection{\textbf{Time Complexity}}
%The multi-head self-attention layer takes $O(hmd'(m+d))$. 
%The time complexity of training process in Fi-GNN consists of three parts. 
%The first part is embed the one-hot feature into field embedding, which is $O(m)$. 
%The second part is receiving neighbors's state information and updating the state of the concerned nodes. each propagation it is $O(m)$ and overall it is $O(mp)$. 
%The last part is estimating the prediction score via two perception networks. 
%According, the total complexity is $O(m(p+1))$.
%Noted that the number of fields ($m$) is usually less than 100.

\subsection{Model Analysis}
\subsubsection{\textbf{Comparison with Previous CTR Models.}}
As discussed before, the previous deep learning based CTR models model high-order interactions in a general paradigm:
raw sparse input multi-filed features are first mapped into dense field embedding vectors, then simply concatenated together and feed into deep neural networks (DNN) or other specifically designed networks to learn high-order feature interactions. 
%However, the unstructured combination of different feature fields will be a bottleneck, which brings difficulty to model sophisticated interactions among the fields in an enough effective and explicit fashion.
The simple unstructured combination of feature fields inevitably limits the capability to model sophisticated interactions among different fields in a sufficiently flexible and explicit fashion.
In this way, the interaction between different fields is conducted in a fixed fashion, no matter how sophisticated the used network is.
In addition, they lack good model explanation.

Since we represent the multi-field features in a graph structure, our proposed model Fi-GNN is able to model interactions among different fields in the form of node interactions.
%As discussed before, we can achieve edge-wise interactions via attentional edge weights and edge-wise transformation function.
Compared with the previous CTR models, Fi-GNN can model the sophisticated feature interaction via flexible edge-wise interaction function, which is more effective and explicit.
Moreover, the edge weights reflecting importance of different interactions can be learnt in Fi-GNN, which provides good model explanations for CTR prediction.   
In fact, if the edge weight is all 1 and the transformation matrix on each edge is same, our model Fi-GNN collapses into FM.
Taking advantage of the great power of GNN, we can apply flexible interactions on different feature fields.

\subsubsection{\textbf{Comparison with Previous GNN Models.}}
Our proposed model Fi-GNN is designed based on GGNN, upon which we mainly make two improvements:
(1) we achieve edge-wise interaction via attentional edge weights and edge-wise transformation;
(2) we introduce an extra residual connection along with GRU to update states, which can help regain the low-order information.

As discussed before, the node interaction on each edge in GNN depends on the edge weight and the transformation function on the edge.
The conventional GGNN uses binary edge weights which fails to reflect the importance of the relations, and a fixed transformation function on all the edges. 
In contrast, our proposed Fi-GNN can model edge-wise interactions via attention edge weights and edge-wise transformation functions.
When the interaction order is high, the node states tend to be smooth, i.e., the states of all the nodes tend to be similar.
The residual connections can help identity the nodes by adding initial node states.

\begin{table}[h]
\centering\caption{Statistics of evaluation datasets.}
\begin{tabular}{cccc} 
\hline
Dataset & \#Instances & \#Fields & \#Features (sparse)  \\
\hline
Criteo & 45,840,617 & 39 & 998,960 \\
Avazu & 40,428,967 & 23 & 1,544,488 \\
%MovieLens-1M & 739,012 & 7 & 3,529 \\
\hline
\end{tabular}\label{tab::dataset}
\end{table}

\begin{table*}
%\large
\centering\caption{Performance Comparison of Different methods. The best performance on each dataset and metric are highlighted. Further analysis is provided in Section \ref{sect:result}.}
\begin{tabular}{llcccccccc} 
\hline
\multirow{2}{*}{Model Type} & \multirow{2}{*}{Model} & \multicolumn{4}{c}{Criteo} & \multicolumn{4}{c}{Avazu}\\
 &  & AUC & RI-AUC & Logloss & RI-Logloss & AUC & RI-AUC & Logloss & RI-Logloss \\
\hline
\multirow{1}{*}{First-order} & LR & 0.7820 & 3.00\% & 0.4695 & 5.43\% & 0.7560 & 2.60\% & 0.3964 & 3.63\% \\
 \hline
\multirow{2}{*}{Second-order} & FM~\cite{rendle2010factorization} & 0.7836 & 2.80\% & 0.4700 & 5.55\% & 0.7706 & 0.72\% & 0.3856 & 0.76\% \\ 
& AFM\cite{xiao2017attentional} & 0.7938 & 1.54\% & 0.4584 & 2.94\% & 0.7718 & 0.57\% & 0.3854 & 0.81\% \\
\midrule
\multirow{5}{*}{High-order} 
& DeepCrossing~\cite{shan2016deep} & 0.8009 & 0.66\% &0.4513 & 1.35\% & 0.7643 & 1.53\% & 0.3889 & 1.67\% \\
& NFM~\cite{he2017neural} & 0.7957 & 1.57\% & 0.4562 & 2.45\% & 0.7708 & 0.70\% & 0.3864 & 1.02\% \\
& CrossNet~\cite{wang2017deep} & 0.7907 & 1.92\% & 0.4591 & 3.10\% & 0.7667 & 1.22\% & 0.3868 & 1.12\% \\
& CIN~\cite{lian2018xdeepfm} & 0.8009 & 0.63\% & 0.4517 & 1.44\% & 0.7758 & 0.05\% & 0.3829 & 0.10\% \\
& Fi-GNN (ours)  & \textbf{0.8062} & 0.00\% & \textbf{0.4453} & 0.00\% & \textbf{0.7762} & 0.00\% & \textbf{0.3825} & 0.00\% \\
\bottomrule
\end{tabular}
\label{tab::results}
\end{table*}

\section{Experiments}
In this section, we conduct extensive experiments to answer the following questions:
\begin{itemize}
\item[\textbf{RQ1}]
 How does our proposed Fi-GNN perform in modeling high-order feature interactions compared with the state-of-the-art models?
\item[\textbf{RQ2}]
 Does our proposed Fi-GNN perform better than original GGNN in modeling high-order feature interactions?
\item[\textbf{RQ3}]
 What are the influences of different model configurations?
 \item[\textbf{RQ4}]
What are the relations between features of different fields? 
Is our proposed model explainable?
\end{itemize}
We first present some fundamental experimental settings before answering these questions.

\subsection{Experiment Setup}

\subsubsection{Datasets}
We evaluate our proposed models on the following two datasets, whose statistics are summarized in Table~\ref{tab::dataset}.

\textbf{1. Criteo\footnote{https://www.kaggle.com/c/criteo-display-ad-challenge}.} This is a famous industry benchmark dataset for CTR prediction, which has 45 million users' click records in 39 anonymous feature fields on displayed ads.
Given a user and the page he is visiting, the goal
is to predict the probability that he will click on a given ad.

\textbf{2. Avazu\footnote{https://www.kaggle.com/c/avazu-ctr-prediction}.} This dataset contains users' click behaviors on displayed mobile ads. 
There are 23 feature fields including user/device features and ad attributes.  
The fields are partial anonymous.

For the two datasets, we remove the infrequent features appearing in less than 10, 5 times respectively and treat them as a single feature ``<unknown>''.
Since the numerical features may have large variance, we normalize numerical values by transforming a value $z$ to $log^2(z)$ if $z > 2$, which is proposed by the winner of Criteo Competition\footnote{\url{https://www.csie.ntu.edu.tw/~r01922136/kaggle-2014-criteo.pdf}}. 
The instances are randomly split in 8:1:1 for training, validation and testing.

\subsubsection{Evaluation Metrics}
We use the following two metrics for model evaluation: AUC (Area Under the ROC curve) and Logloss (cross entropy).

\textbf{AUC} measures the probability that a positive instance will be ranked higher than a randomly chosen negative one.  
A higher AUC indicates a better performance.

\textbf{Logloss} measures the distance between the predicted score and the true label for each instance.
A lower Logloss indicates a better performance.

\textbf{Relative Improvement (RI)}. It should be noted that a small improvement with respect to AUC is regarded significant for real-world CTR tasks \cite{cheng2016wide,guo2017deepfm,wang2017deep,lian2018xdeepfm}. 
In order to estimate the relative improvement of our model achieves over the compared models, we here measure \textbf{RI-AUC} and \textbf{RI-Logloss}, which can be formulated as,
\begin{equation}
\textit{RI}\text{-}\textit{X} = \dfrac {\left |\textit{X}(model)-\textit{X}(base) \right |}{\textit{X}(base)} *100\%~,
\end{equation}
where $\left | x \right |$ returns the absolute value of x, $X$ can be either AUC or Logloss, $\textit{model}$ refers to our proposed model and $\textit{base}$ refers to the compared model.

%It should be noted that the previous improvements with respect to AUC are usually at \textit{\textbf{0.001-level}}, which is regarded significant for CTR prediction task \cite{cheng2016wide,guo2017deepfm,wang2017deep,lian2018xdeepfm}.

\subsubsection{Baselines} 
As described in Section \ref{sect:related}, 
%the previous methods can  (A) the linear approach that model first-order feature interaction. (B) factorization machines-based methods that take into account second-order combinatorial features. (C) techniques that can capture high-order feature interactions.
the early approaches can be categorized into three types: 
(A) Logistic Regression (LR) which models first-order interaction;
(B) Factorization Machine (FM) based linear models which model second-order interactions; 
(C) Deep learning based models which model high-order interactions on the concatenated field embedding vectors. 

We select the following representative methods of three types to compare with ours.

\textbf{LR} (A) models first-order interaction on the linear combination of raw individual features.  

\textbf{FM}~\cite{rendle2010factorization} (B) models second-order feature interactions from vector inner products. 

\textbf{AFM}~\cite{xiao2017attentional} (B) is a extent of FM, which considers the weight of different second-order feature interactions by using attention mechanism.
It is one of the state-of-the-art models that model second-order feature interactions. 

\textbf{DeepCrossing}~\cite{shan2016deep} (C) utilizes DNN with residual connections to learn high-order feature interactions in an implicit fashion.
%\textbf{Deep\&Cross}~\cite{wang2017deep} (C). Deep\&Cross is the state-of-the-art model that learns high-order feature interactions. Specifically, a cross network, which takes the outer product of feature vector at bit-wise level, is proposed to model feature interactions explicitly, and combines with a feed-forward neural network to model feature interactions implicitly.

\textbf{NFM}~\cite{he2017neural} (C) utilizes a Bi-Interaction Pooling layer to model the second-order interactions, and then feeds the concatenated second-order combinatorial features into DNNs to model high-order interactions.

\textbf{CrossNet (Deep\&Cross) }~\cite{wang2017deep} (C) is the core of Deep\&Cross model, which tries to model feature interactions explicitly by taking outer product of concatenated feature vector at the bit-wise level.

\textbf{CIN (xDeepFM)}~\cite{lian2018xdeepfm} (C) is the core of xDeepFM model, which takes outer product of stacked feature matrix at vector-wise level.

%We will compare with the full models of CrossNet and CIN, i.e., Deep\&Cross and xDeepFM, in a joint training setting later.  

%\textbf{HOFM}~\cite{blondel2016higher} (C). HOFM proposes efficient kernel-based algorithms for training high-order factorization machines. Follow settings in~\citeauthor{blondel2016higher}~\cite{blondel2016higher} and \citeauthor{he2017neural}~\cite{he2017neural}, we build a third-order factorization machine using public implementation\footnote{https://github.com/geffy/tffm}.

\subsubsection{Implementation Details}
We implement our method using Tensorflow\footnote{The code is released at \url{https://github.com/CRIPAC-DIG/Fi_GNN}}. The optimal hyper-parameters are determined by the grid search strategy. 
%We experimentally find that the model achieves optimal performance with the learning rate as 0.001, batch size as 16, β as 0.2, λ as 0.001, d as 12 and T as 3.  
Implementation of baselines follows \cite{song2018autoint}.
Dimension of field embedding vectors is 16 and batch size is 1024 for all methods. 
DeepCrossing has four feed-forward layers, each with 100 hidden units.  
NFM has one hidden layer of size 200 on top of Bi-Interaction layer as recommended in the paper \cite{he2017neural}. 
There are three interaction layers for both CrossNet and CIN. 
%AutoInt has two attention heads in each layer.
All the experiments were conducted over a sever equipped with 8 NVIDIA Titan X GPUs.

\begin{figure*}[hbtp]
%\centering
\subfigure[edge-wise interaction (E) and residual connections (R)]{
\begin{minipage}[b]{0.5\textwidth}
%\centering
\label{fig:ablation_er} %% label for first subfigure
\includegraphics[width=1\textwidth]{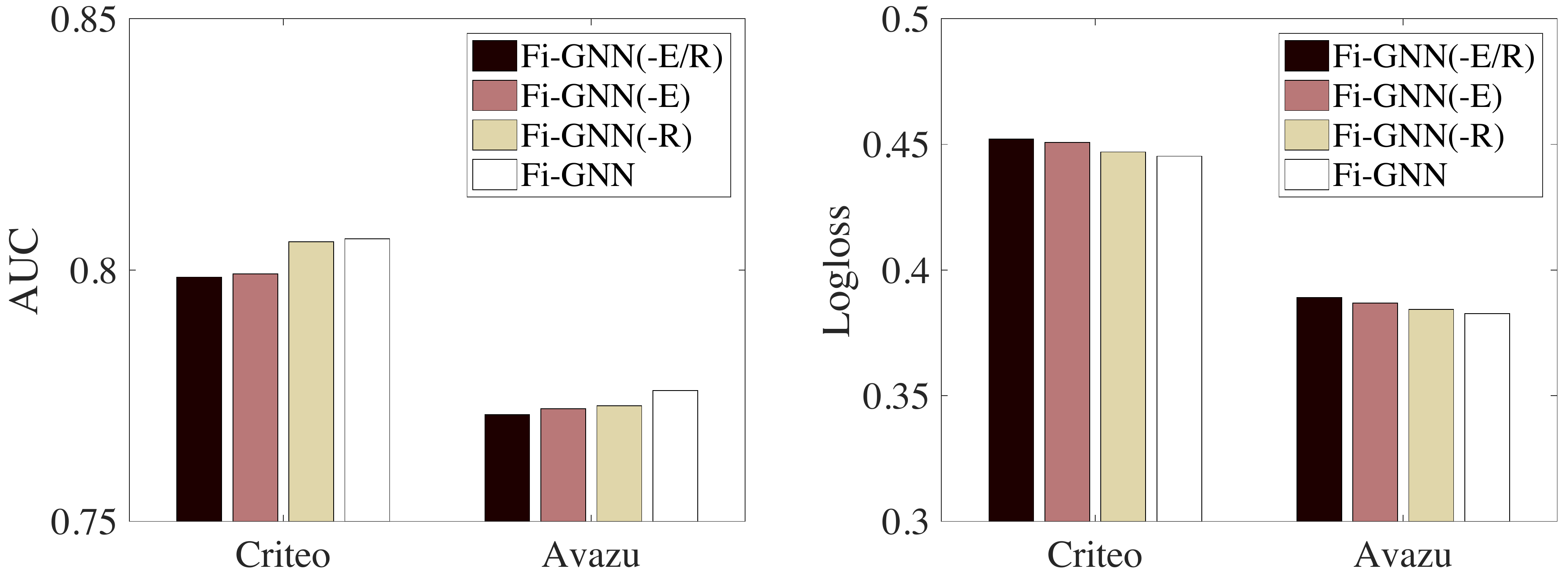}

\end{minipage}%
}%
\subfigure[attentional edge weight (W) and edge-wise transformation (T)]{
\begin{minipage}[b]{0.5\textwidth}
%\centering
\label{fig:ablation_wt} %% label for first subfigure
\includegraphics[width=1\textwidth]{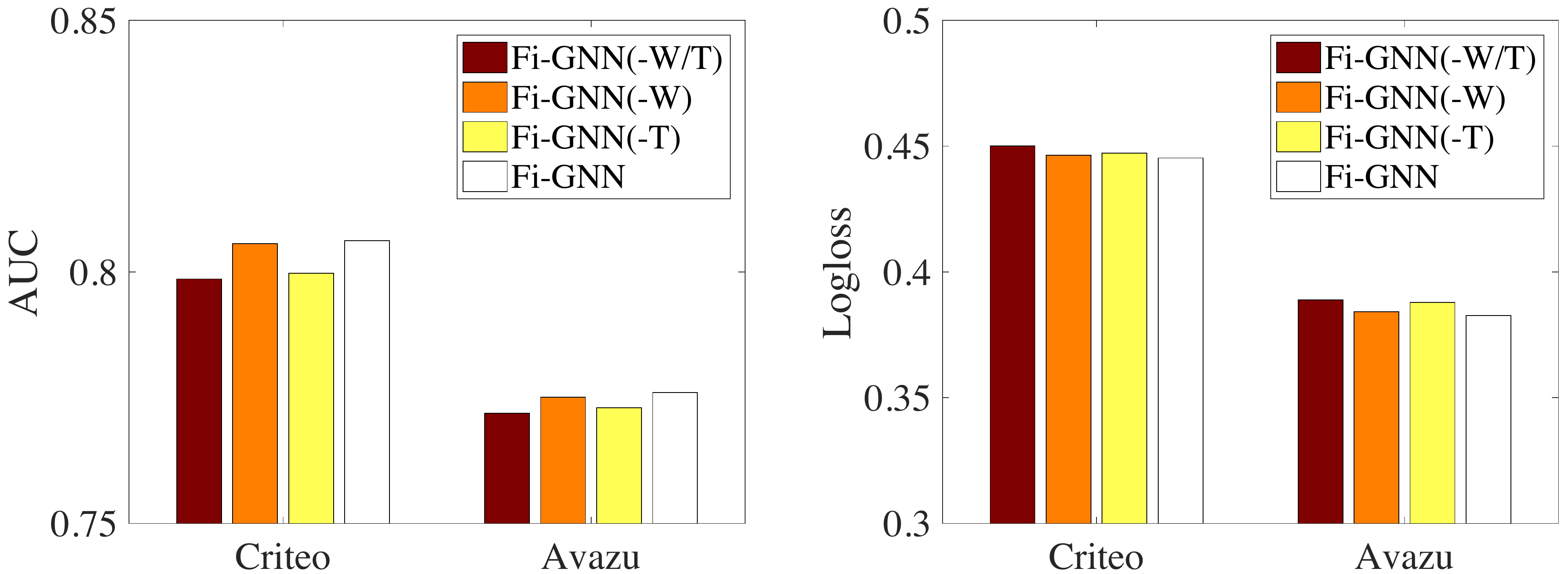}
\end{minipage}%
}%
\caption{Two groups of ablation studies on Fi-GNN.}
\label{fig:performance}
\end{figure*}

\subsection{Model Comparison (RQ1)}\label{sect:result}
The performance of different methods is summarized in Table \ref{tab::results}, from which we can obtain the following observations:
 
\begin{itemize}
\item[(1)]
LR achieves the worst performance among these baselines, which proves that the individual features is insufficient in CTR prediction.
\item[(2)]
FM and AFM, which model second-order feature interactions, outperform LR on all datasets, indicating that it's effective to model pair-wise interaction between feature fields. 
In addition, AFM achieves better performance than FM, which proves the effectiveness of attention on different interactions.
\item[(3)]
The methods modeling high-order interaction mostly outperform the methods that model second-order interactions.
This indicates the second-order feature interactions is not sufficient. 
\item[(4)]
DeepCrossing outperforms NFM, proving the effectiveness of residual connections in CTR prediction.
\item[(5)]
Our proposed Fi-GNN achieves best performance among all these methods on two datasets.
Considering the fact that previous improvements with respect to AUC at \textbf{0.001-level} are regarded significant for CTR prediction task, our proposed method shows great superiority over these state-of-the-arts especially on Criteo dataset, owing to the great representative power of graph structure and the effectiveness of GNN on modeling node interactions.
\item[(6)] Compared with these baselines, the relative improvement of our model achieves on Criteo dataset is higher than that on Avazu dataset. This might be attributed to that there are more feature fields in Criteo dataset, which can take more advantage of the representative power of graph structure.
\end{itemize}

%(1) LR achieves the worst performance among these baselines, which proves that the individual features is insufficient in CTR prediction.
%
%(2) FM and AFM, which model second-order feature interactions, outperform LR on all datasets, indicating that it's effective to model pair-wise interaction on the combinatorial features.
%
%(3) NFM and DeepCrossing outperform the methods that model second-order interactions while CrossNet not. 
%This indicates the inferiority of some models that capture high-order feature interactions, which may due to the fact that they learn feature interactions in an implicit fashion.
%
%(4) DeepCrossing outperforms NFM, proving the effectiveness of residual connections in CTR prediction.
%
%(5) Our proposed Fi-GNN achieves best performance on all the three datasets and both the metrics.
%Considering the fact that previous improvements with respect to AUC at \textbf{0.001-level} are regarded significant for CTR prediction task,
%our proposed method shows great superiority over these state-of-the-arts, which may owe to the great representative power of graph structure and the effectiveness of GNN on handling graph-structured data.
            
\subsection{Ablation Study (RQ2)}\label{sect:comp_gnn}

Our proposed model Fi-GNN is based on GGNN, upon which we mainly make two improvements:
(1) we achieve edge-wise node interactions via attentional edge weights and edge-wise transformation;
(2) we introduce extra residual connections to update state along with GRU.
To evaluate the effectiveness of the two improvements on modeling node interactions, we conduct ablation study and compare the following three variants of Fi-GNN:

\textbf{Fi-GNN(-E/R)}:
Fi-GNN without the two above mentioned improvements: edge-wise node interactions (\textbf{E}) and residual connections (\textbf{R}).

\textbf{Fi-GNN(-E)}:
Fi-GNN without edge-wise interactions (\textbf{E}).

\textbf{Fi-GNN(-R)}:
Fi-GNN without residual connections (\textbf{R}), which is also GGNN with edge-wise interactions. 

The performance comparison is shown in Figure \ref{fig:ablation_er}, from which we can obtain the following observations:
\begin{itemize}
\item[(1)]
%Fi-GNN(-E) and Fi-GNN(-R) both outperform GGNN, which proves the effectiveness of residual connections and edge-wise interaction in modeling node interactions.
%Fi-GNN(-R) outperform Fi-GNN(-E/R) by a margin, which suggests that it is crucial to model the edge-wise interaction.
%Fi-GNN(-E) also achieves better performance than Fi-GNN(-E/R), proving the effectiveness of residual connections. 
%The full model Fi-GNN outperforms the three variants, which proves that the effectiveness of residual connections and edge-wise interaction in modeling node interactions.
Compared with FiGNN，the performance of Fi-GNN(-E) drops by a large margin, suggesting that it's crucial to model the edge-wise interaction.
Fi-GNN(-E) achieves better performance than Fi-GNN(-E/R), proving that the residual connections can indeed provide useful information. 
\item[(2)]
The full model Fi-GNN outperforms the three variants, indicating that the two improvements we make, i.e., residual connections and edge-wise interactions, can jointly boost the performance.
\end{itemize}

%(1) Fi-GNN$_{\text{R}}$ outperforms GGNN, which proves the effectiveness of residual connections in modeling high-order feature interactions; 
%       
%(2) Fi-GNN$_{\text{E}}$ outperforms GGNN, demonstrating the effectiveness of edge-wise interaction in modeling node interactions; 
%   
%(3) Fi-GNN achieve best performance among these variants, indicating that the two improvements we make, i.e., residual connections and edge-wise interactions, can jointly boost the performance.

We take two measures to achieve edge-wise node interactions in Fi-GNN: attentional edge weight (\textbf{W}) and edge-wise transformation (\textbf{T}).
To further investigate where dose the great improvement come from, we conduct another ablation study and compare the following three variants of Fi-GNN:

\textbf{Fi-GNN(-W/T)}: Fi-GNN without self-adaptive adjacency matrix (\textbf{W}) and edge-wise transformation (\textbf{T}), i.e., uses binary adjacency matrix (all the edge weights are 1) and a shared transformation matrix on all the edges.
It is also \textbf{Fi-GNN-(E)},

\textbf{Fi-GNN(-W)}: FI-GNN without attentional edge weights, i.e., uses binary adjacency matrix.

\textbf{Fi-GNN(-T)}: FI-GNN without edge-wise transformation,
i.e., uses a shared transformation on all the edges. 

The performance comparison is shown in Figure \ref{fig:ablation_er}.
We can see that Fi-GNN(-T) and Fi-GNN(-W) both outperform Fi-GNN(-W/T), which proves their effectiveness.
Nevertheless, Fi-GNN(-W) achieves greater improvements than Fi-GNN(-T), suggesting that the edge-wise transformation is more effective than attentional edge weights in modeling edge-wise interaction.
This is quite reasonable since the transformation matrix oughts to have stronger influence on interactions than a scalar attentional edge weight.
In addition, Fi-GNN achieves the best performance demonstrates that it's crucial to take both the two measures to model edge-wise interaction.

%\begin{itemize}
%\item[(1)] Fi-GNN(-T) and Fi-GNN(-W) both outperform Fi-GNN(-W/T), proving the effectiveness of attentional edge weights and edge-wise transformation.
%\item[(2)]
%Fi-GNN achieves best performance among these variants, indicating that the two improvements we make, i.e., residual connections and edge-wise interactions, can jointly boost the performance.
%\end{itemize}

%\begin{table}
%\centering
%\caption{Ablation study on attentional edge weight (\textbf{W}) and edge-wise transformation (\textbf{T}).} 
%%The experiments are conducted on Criteo and Avazu datasets.}       
%
%
%\begin{tabularx}{1.0\linewidth}{l>{\centering\arraybackslash}l>{\centering\arraybackslash}X>{\centering\arraybackslash}X}
%\toprule
%Models & Datasets & AUC & Logloss \\
%\midrule
%\multirow{2}{*}{Fi-GNN(-W/T)} 
%& Criteo & 0.7911 & 0.4554 \\
%& Avazu & 0.8002 & 0.3867 \\
%\midrule
%\multirow{2}{*}{Fi-GNN(-W)} 
%& Criteo & 0.8061 & 0.4434 \\
%& Avazu & 0.8114 & 0.3819 \\
%\midrule
%\multirow{2}{*}{Fi-GNN(-T)} 
%& Criteo & 0.7989 & 0.4484 \\
%& Avazu & 0.8033 & 0.3848 \\
% \midrule
%\multirow{2}{*}{Fi-GNN} 
%& Criteo & 0.8082 & 0.4411 \\
%& Avazu & 0.8120 & 0.3817 \\
%\bottomrule
%\end{tabularx}\label{tab:ablation2}
%\vspace{-2mm}
%\end{table}

\begin{figure}[t]
\centering
\subfigure[State Dimensionality]{
\begin{minipage}[b]{0.24\textwidth}
%\centering
\label{fig:hidden} %% label for first subfigure
\includegraphics[width=1\textwidth]{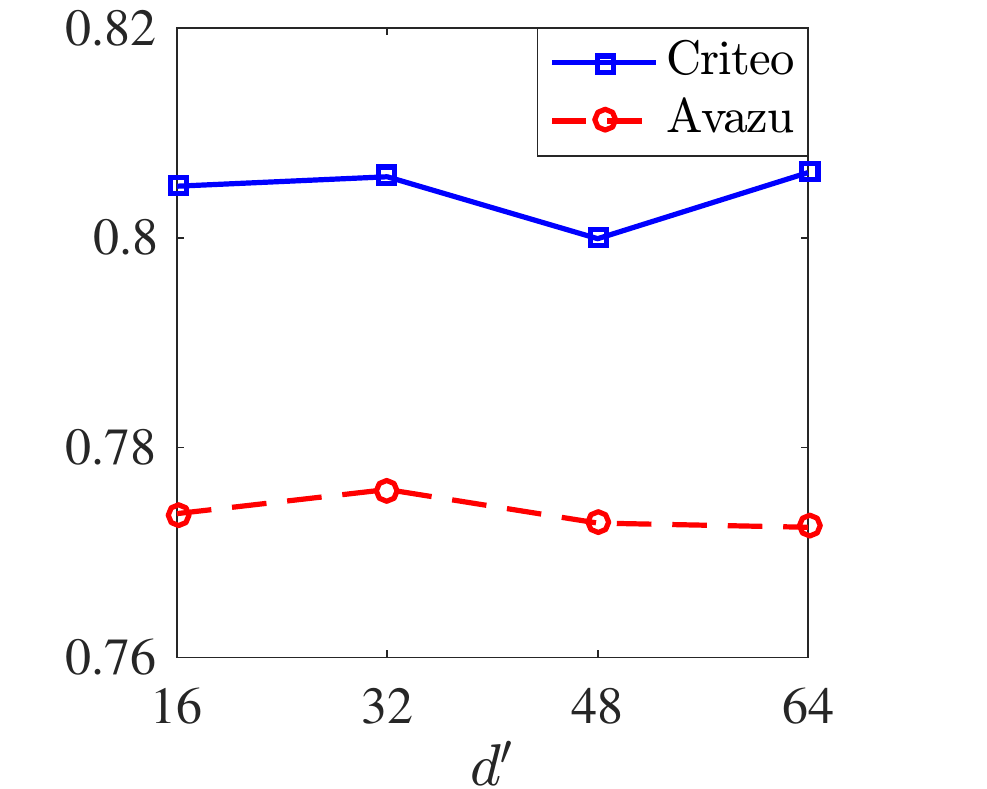}

\end{minipage}%
}%
\subfigure[Interaction Step]{
\begin{minipage}[b]{0.24\textwidth}
%\centering
\label{fig:order} %% label for first subfigure
\includegraphics[width=1\textwidth]{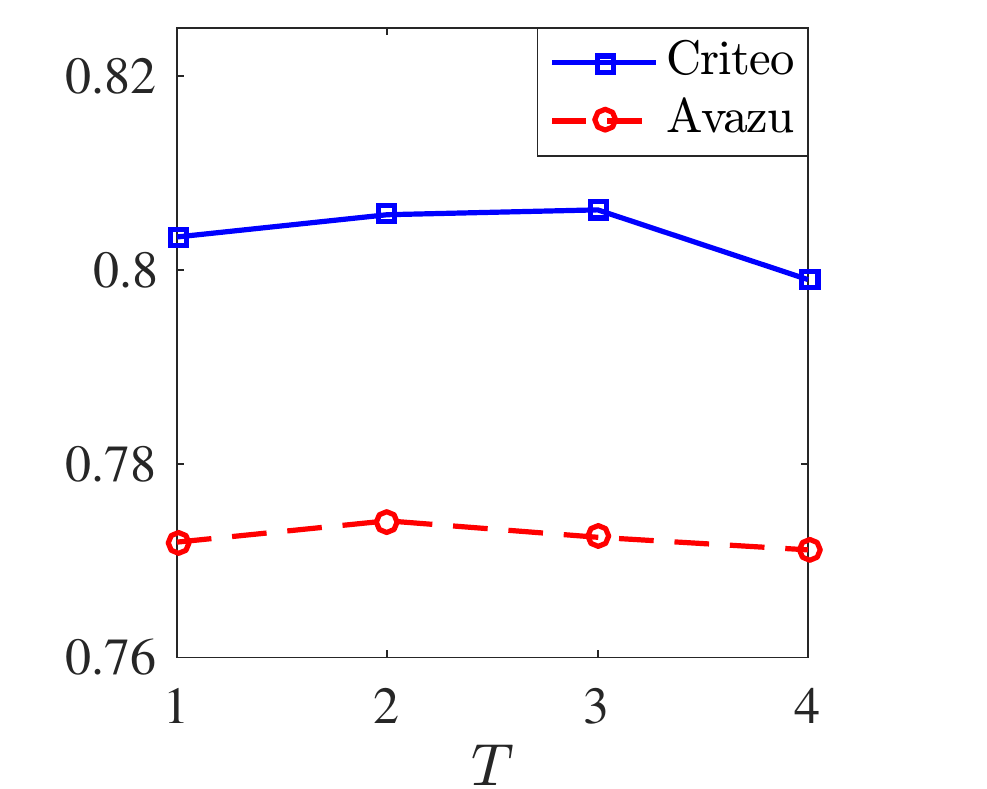}
\end{minipage}%
}%
\caption{AUC performance with different state dimensionality $D$ (left) and interaction step $T$ (right) on Criteo and Avazu dataset.}
\label{fig:performance}
\vspace{-4mm}
\end{figure}

\subsection{Hyper-Parameter Study (RQ3)}

\subsubsection{\textbf{Influence of different state dimensionality.}}
We first investigate how the performance changes w.r.t. the dimension of the node states $d'$, which is also the output size of the initial multi-head self-attention layer. 
The results on Criteo and Avazu datasets are shown in Figure \ref{fig:hidden}.
On Avazu dataset, the performance first increases and then begins to decrease when the dimension size reaches 32, which indicates that state size of 32 has been represented enough information and the model is overfitted when too many parameters are used. 
Nevertheless, on Criteo dataset, the performance peaks with the dimension size of 64, which is reasonable since the dataset is more complexed which needs larger dimension size to carry out enough information.
%On both datasets, we can see that the performance continuously increase along with the increasing of the dimension size since larger size can provide sufficient information. 
%When the dimension size reaches 16, the performances begin to decrease, which may due to that state of this size has been represented enough information and the model is overfitted when too many parameters are used. 

\subsubsection{\textbf{Influence of different interaction steps.}}
We are interested in what the optimal highest order of feature interactions is.  
Our proposed Fi-GNN can answer the question, since the interaction step $T$ equals to the highest order of feature interaction.
Therefore, we conduct experiments on how the performance changes w.r.t. the highest order of feature interaction, i.e., the interaction step $T$.
The results on Criteo and Avazu datasets are shown in Figure \ref{fig:order}.
On Avazu datasets, we can see that the performance increases along with the increasing of $T$ until it reaches 2, after that the performance starts to decrease.
By contrast, the performance peaks when $T=3$ on Criteo dataset.
This finding suggests 2-order and 3-order interactions are enough for Avazu and Criteo dataset, respectively.
It is reasonable since the Avazu and Criteo datasets have 23 and 39 feature fields, respectively.
Thus the Criteo dataset needs more interaction steps for the field nodes to fully interact with other nodes in the feature graphs.
%The optimal order of interaction depends on the number of feature fields.
%Criteo and Avazu datasets have 39 and 23 feature fields respectively, 3-order interaction is sufficient for the nodes to be aware of the  

\subsection{Model Explanation (RQ4)} \label{sect:explannation}
In this section, we will answer the question that can Fi-GNN provide explanations.
We apply attention mechanisms on the edges and nodes in the feature graphs and obtain attentional edge weights and attentional node weights respectively, which can provide explanations from different aspects.

\begin{figure}[t]
\centering
\includegraphics[width=1\linewidth]{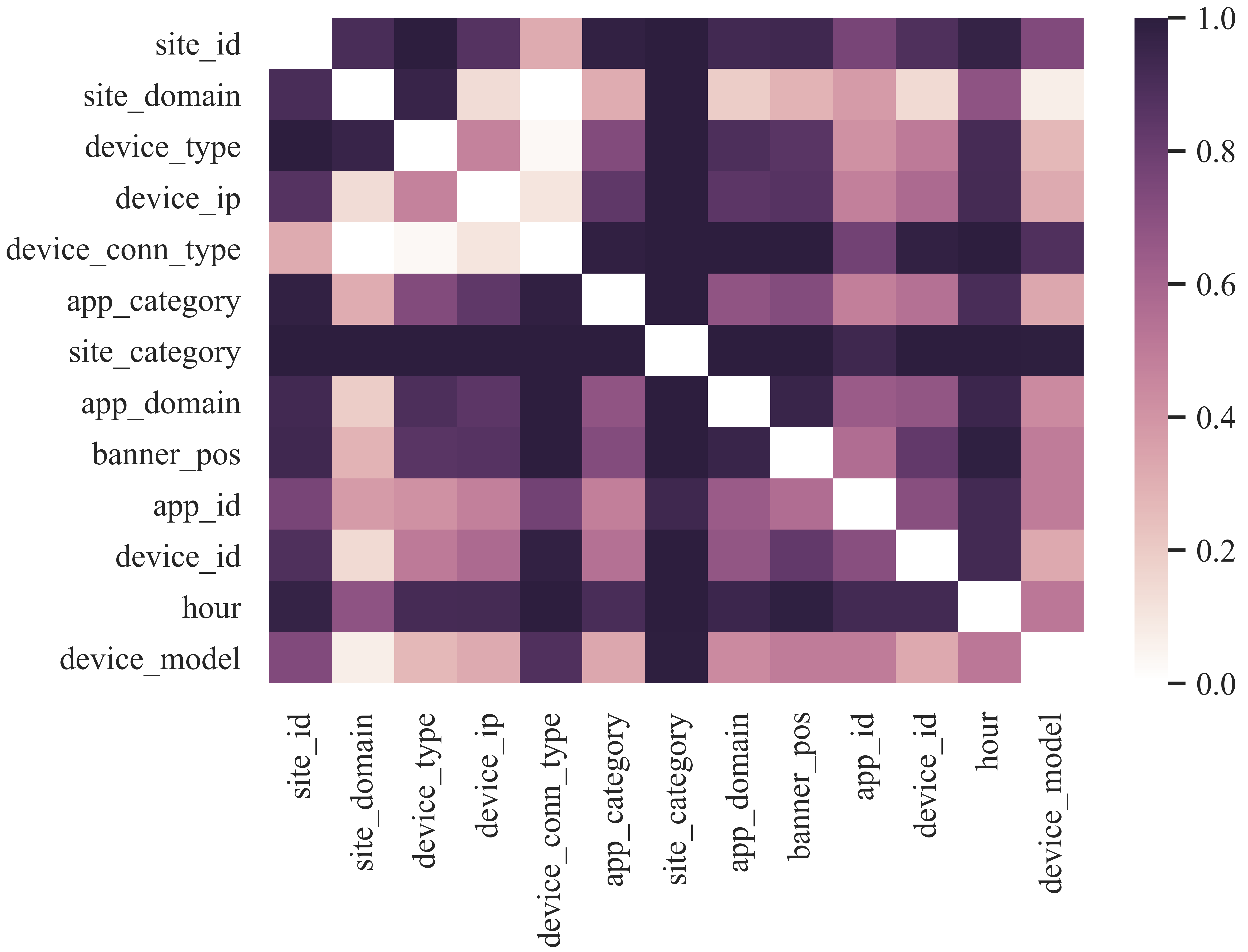}
\caption{Heat map of attentional edge weights at the global-level on Avazu, which reflects the importance of relations between different feature fields.}
%The input could be image, text or any other modality of items.}
\label{fig:heatmap_edge}
\end{figure}

\subsubsection{\textbf{Attentional Edge weights.}}
The attentional edge weight reflects the importance of interaction between the two connected field nodes, which can also reflect the relation of the two feature fields.
Higher the weight is, stronger the relation is.
Figure \ref{fig:heatmap_edge} presents the heat map of the globally averaged adjacency matrix of all the samples in Avazu dataset, which can reflect the relations between different fields in a global level. 
Since they are some anonymous feature fields, we only show the remaining 13 feature fields with real meanings.

As can be seen, some feature fields tend to have a strong relations with others, such as \textsf{site\_category} and \textsf{site\_id}.
This makes sense since the two feature field both corresponds to the website where the impressions are put on. They contain the main contextual information of impressions. 
%The website a user surfs on also imply its present appetite.
\textsf{Hour} is another feature which have close relations with others. It is reasonable since Avazu focuses on mobile scene, where user surfing online at any time of a day.  
The surfing time has strong influence on other advertising features.  
On the other hand, \textsf{device\_ip} and \textsf{device\_id} seem to have weak relations with other feature fields.
This may due to that they nearly equal to user identity, which is relatively fixed and hard to be influenced by other features. 

\begin{figure}[t]
\centering
\includegraphics[width=1\linewidth]{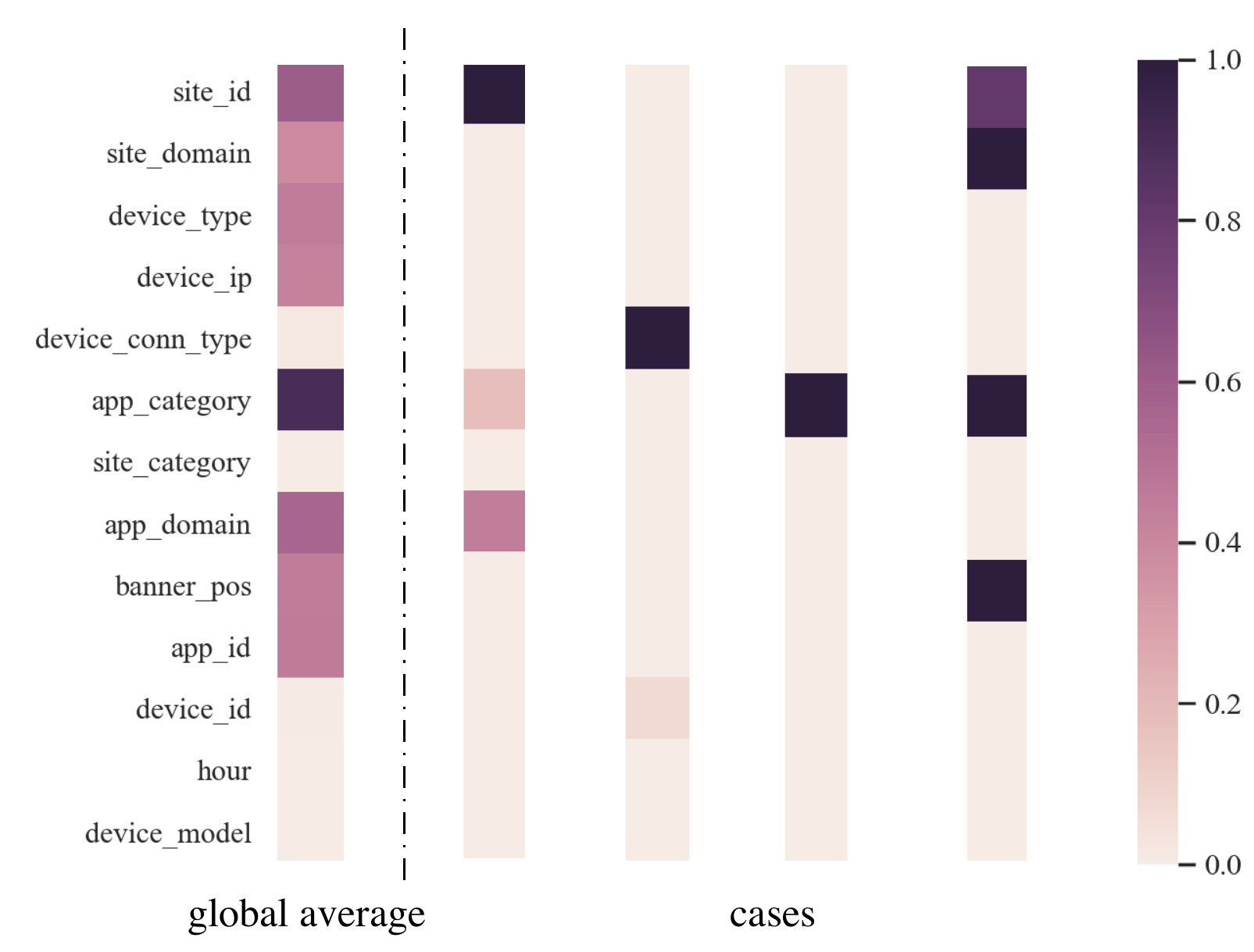}
\caption{Heat map of attentional node weights at both global- and case-level on Avazu, which reflects the importance of different feature fields on the final prediction.}
%The input could be image, text or any other modality of items.}
\label{fig:heatmap_node}
\vspace{-5mm}
\end{figure}

\subsubsection{\textbf{Attentional Node weights.}}
The attentional node weights reflect the importances of feature fields' influence on the overall prediction score.
Figure \ref{fig:heatmap_node} presents the heat map of global-level and case-level attentional node weights.
The leftmost is an globally averaged one of all the samples in Avazu dataset.
The left four are randomly selected, whose predicted scores are $[0.97, 0.12, 0.91, 0.99]$, and labels are $[1, 0, 1, 1]$ respectively. 
At the global level, we can see that the feature field \textsf{app\_category} have the strongest influence on the clicking behaviors. 
It is reasonable since Avazu focuses on mobile scene, where the app is the most important factor. 
At the case level, we observe that the final clicking behavior mainly depends on one critical feature field in most cases.

\section{Conclusions}
In this paper, we point out the limitations of the previous CTR models which consider multi-field features as an unstructured combination of feature fields.
To overcome these limitations, we propose to represent the multi-field features in a graph structure for the first time, where each node corresponds to a feature field and different fields can interact through edges.
Therefore, modeling feature interactions can be converted to modeling node interaction on the graph.  
%\item We design a GNN-based model Fi-GNN to model certain order feature interactions on the graph-structured features in an explicit, vector-wise fashion.
To this end, we design a novel model Fi-GNN which is able to model sophisticated interactions among feature fields in a flexible and explicit fashion.
Overall, we propose a new paradigm of CTR prediction: represent multi-field features in a graph structure and convert the task of modeling feature interactions to modeling node interactions on graphs, which may motivate the future work in this line.

% The acknowledgments section is defined using the "acks" environment (and NOT an unnumbered section). This ensures
% the proper identification of the section in the article metadata, and the consistent spelling of the heading.

\begin{acks}
This work is supported by National Natural Science Foundation of China (61772528, 61871378) and National Key Research and Development Program (2016YFB1001000, 2018YFB1402600).
\end{acks}

% The next two lines define the bibliography style to be used, and the bibliography file.
\bibliographystyle{ACM-Reference-Format}
\bibliography{sample-base}


\begin{thebibliography}{36}

%%% the Web address.

\ifx \showCODEN    \undefined \def \showCODEN     #1{\unskip}     \fi
\ifx \showDOI      \undefined \def \showDOI       #1{#1}\fi
\ifx \showISBNx    \undefined \def \showISBNx     #1{\unskip}     \fi
\ifx \showISBNxiii \undefined \def \showISBNxiii  #1{\unskip}     \fi
\ifx \showISSN     \undefined \def \showISSN      #1{\unskip}     \fi
\ifx \showLCCN     \undefined \def \showLCCN      #1{\unskip}     \fi
\ifx \shownote     \undefined \def \shownote      #1{#1}          \fi
\ifx \showarticletitle \undefined \def \showarticletitle #1{#1}   \fi
\ifx \showURL      \undefined \def \showURL       {\relax}        \fi
% invisible to TeX
\providecommand\bibfield[2]{#2}
\providecommand\bibinfo[2]{#2}
\providecommand\natexlab[1]{#1}
\providecommand\showeprint[2][]{arXiv:#2}

\bibitem[\protect\citeauthoryear{Beck, Haffari, and Cohn}{Beck
  et~al\mbox{.}}{2018}]%
        {beck2018graph}
\bibfield{author}{\bibinfo{person}{Daniel Beck}, \bibinfo{person}{Gholamreza
  Haffari}, {and} \bibinfo{person}{Trevor Cohn}.}
  \bibinfo{year}{2018}\natexlab{}.
\newblock \showarticletitle{Graph-to-sequence learning using gated graph neural
  networks}.
\newblock \bibinfo{journal}{\emph{arXiv preprint arXiv:1806.09835}}
  (\bibinfo{year}{2018}).
\newblock


\bibitem[\protect\citeauthoryear{Cheng, Koc, Harmsen, Shaked, Chandra, Aradhye,
  Anderson, Corrado, Chai, Ispir, et~al\mbox{.}}{Cheng et~al\mbox{.}}{2016}]%
        {cheng2016wide}
\bibfield{author}{\bibinfo{person}{Heng-Tze Cheng}, \bibinfo{person}{Levent
  Koc}, \bibinfo{person}{Jeremiah Harmsen}, \bibinfo{person}{Tal Shaked},
  \bibinfo{person}{Tushar Chandra}, \bibinfo{person}{Hrishi Aradhye},
  \bibinfo{person}{Glen Anderson}, \bibinfo{person}{Greg Corrado},
  \bibinfo{person}{Wei Chai}, \bibinfo{person}{Mustafa Ispir}, {et~al\mbox{.}}}
  \bibinfo{year}{2016}\natexlab{}.
\newblock \showarticletitle{Wide \& deep learning for recommender systems}. In
  \bibinfo{booktitle}{\emph{Proceedings of the 1st workshop on deep learning
  for recommender systems}}. ACM, \bibinfo{pages}{7--10}.
\newblock


\bibitem[\protect\citeauthoryear{Cho, Van~Merri{\"e}nboer, Gulcehre, Bahdanau,
  Bougares, Schwenk, and Bengio}{Cho et~al\mbox{.}}{2014}]%
        {cho2014learning}
\bibfield{author}{\bibinfo{person}{Kyunghyun Cho}, \bibinfo{person}{Bart
  Van~Merri{\"e}nboer}, \bibinfo{person}{Caglar Gulcehre},
  \bibinfo{person}{Dzmitry Bahdanau}, \bibinfo{person}{Fethi Bougares},
  \bibinfo{person}{Holger Schwenk}, {and} \bibinfo{person}{Yoshua Bengio}.}
  \bibinfo{year}{2014}\natexlab{}.
\newblock \showarticletitle{Learning phrase representations using RNN
  encoder-decoder for statistical machine translation}.
\newblock \bibinfo{journal}{\emph{arXiv preprint arXiv:1406.1078}}
  (\bibinfo{year}{2014}).
\newblock


\bibitem[\protect\citeauthoryear{Cui, Li, Wu, Zhang, and Wang}{Cui
  et~al\mbox{.}}{2019}]%
        {cui2019dressing}
\bibfield{author}{\bibinfo{person}{Zeyu Cui}, \bibinfo{person}{Zekun Li},
  \bibinfo{person}{Shu Wu}, \bibinfo{person}{Xiaoyu Zhang}, {and}
  \bibinfo{person}{Liang Wang}.} \bibinfo{year}{2019}\natexlab{}.
\newblock \showarticletitle{Dressing as a Whole: Outfit Compatibility Learning
  Based on Node-wise Graph Neural Networks}.
\newblock \bibinfo{journal}{\emph{arXiv preprint arXiv:1902.08009}}
  (\bibinfo{year}{2019}).
\newblock


\bibitem[\protect\citeauthoryear{Grover and Leskovec}{Grover and
  Leskovec}{2016}]%
        {Grover2016node2vec}
\bibfield{author}{\bibinfo{person}{Aditya Grover} {and} \bibinfo{person}{Jure
  Leskovec}.} \bibinfo{year}{2016}\natexlab{}.
\newblock \showarticletitle{node2vec: Scalable feature learning for networks}.
  In \bibinfo{booktitle}{\emph{Proceedings of the 22nd ACM SIGKDD international
  conference on Knowledge discovery and data mining}}. ACM,
  \bibinfo{pages}{855--864}.
\newblock


\bibitem[\protect\citeauthoryear{Guo, Tang, Ye, Li, and He}{Guo
  et~al\mbox{.}}{2017}]%
        {guo2017deepfm}
\bibfield{author}{\bibinfo{person}{Huifeng Guo}, \bibinfo{person}{Ruiming
  Tang}, \bibinfo{person}{Yunming Ye}, \bibinfo{person}{Zhenguo Li}, {and}
  \bibinfo{person}{Xiuqiang He}.} \bibinfo{year}{2017}\natexlab{}.
\newblock \showarticletitle{DeepFM: a factorization-machine based neural
  network for CTR prediction}. In \bibinfo{booktitle}{\emph{Proceedings of the
  26th International Joint Conference on Artificial Intelligence}}. AAAI Press,
  \bibinfo{pages}{1725--1731}.
\newblock


\bibitem[\protect\citeauthoryear{Hamilton, Ying, and Leskovec}{Hamilton
  et~al\mbox{.}}{2017}]%
        {hamilton2017inductive}
\bibfield{author}{\bibinfo{person}{Will Hamilton}, \bibinfo{person}{Zhitao
  Ying}, {and} \bibinfo{person}{Jure Leskovec}.}
  \bibinfo{year}{2017}\natexlab{}.
\newblock \showarticletitle{Inductive representation learning on large graphs}.
  In \bibinfo{booktitle}{\emph{Advances in Neural Information Processing
  Systems}}. \bibinfo{pages}{1024--1034}.
\newblock


\bibitem[\protect\citeauthoryear{He and Chua}{He and Chua}{2017}]%
        {he2017neural}
\bibfield{author}{\bibinfo{person}{Xiangnan He} {and} \bibinfo{person}{Tat-Seng
  Chua}.} \bibinfo{year}{2017}\natexlab{}.
\newblock \showarticletitle{Neural factorization machines for sparse predictive
  analytics}. In \bibinfo{booktitle}{\emph{Proceedings of the 40th
  International ACM SIGIR conference on Research and Development in Information
  Retrieval}}. ACM, \bibinfo{pages}{355--364}.
\newblock


\bibitem[\protect\citeauthoryear{Juan, Zhuang, Chin, and Lin}{Juan
  et~al\mbox{.}}{2016}]%
        {juan2016field}
\bibfield{author}{\bibinfo{person}{Yuchin Juan}, \bibinfo{person}{Yong Zhuang},
  \bibinfo{person}{Wei-Sheng Chin}, {and} \bibinfo{person}{Chih-Jen Lin}.}
  \bibinfo{year}{2016}\natexlab{}.
\newblock \showarticletitle{Field-aware factorization machines for CTR
  prediction}. In \bibinfo{booktitle}{\emph{Proceedings of the 10th ACM
  Conference on Recommender Systems}}. ACM, \bibinfo{pages}{43--50}.
\newblock


\bibitem[\protect\citeauthoryear{Kipf and Welling}{Kipf and Welling}{2016}]%
        {kipf2016semi}
\bibfield{author}{\bibinfo{person}{Thomas~N Kipf} {and} \bibinfo{person}{Max
  Welling}.} \bibinfo{year}{2016}\natexlab{}.
\newblock \showarticletitle{Semi-supervised classification with graph
  convolutional networks}.
\newblock \bibinfo{journal}{\emph{arXiv preprint arXiv:1609.02907}}
  (\bibinfo{year}{2016}).
\newblock


\bibitem[\protect\citeauthoryear{Li, Tapaswi, Liao, Jia, Urtasun, and
  Fidler}{Li et~al\mbox{.}}{2017}]%
        {li2017situation}
\bibfield{author}{\bibinfo{person}{Ruiyu Li}, \bibinfo{person}{Makarand
  Tapaswi}, \bibinfo{person}{Renjie Liao}, \bibinfo{person}{Jiaya Jia},
  \bibinfo{person}{Raquel Urtasun}, {and} \bibinfo{person}{Sanja Fidler}.}
  \bibinfo{year}{2017}\natexlab{}.
\newblock \showarticletitle{Situation recognition with graph neural networks}.
  In \bibinfo{booktitle}{\emph{Proceedings of the IEEE International Conference
  on Computer Vision}}. \bibinfo{pages}{4173--4182}.
\newblock


\bibitem[\protect\citeauthoryear{Li, Tarlow, Brockschmidt, and Zemel}{Li
  et~al\mbox{.}}{2015}]%
        {li2015gated}
\bibfield{author}{\bibinfo{person}{Yujia Li}, \bibinfo{person}{Daniel Tarlow},
  \bibinfo{person}{Marc Brockschmidt}, {and} \bibinfo{person}{Richard Zemel}.}
  \bibinfo{year}{2015}\natexlab{}.
\newblock \showarticletitle{Gated graph sequence neural networks}.
\newblock \bibinfo{journal}{\emph{arXiv preprint arXiv:1511.05493}}
  (\bibinfo{year}{2015}).
\newblock


\bibitem[\protect\citeauthoryear{Li, Cui, Wu, Zhang, and Wang}{Li
  et~al\mbox{.}}{2019}]%
        {li2019semi}
\bibfield{author}{\bibinfo{person}{Zekun Li}, \bibinfo{person}{Zeyu Cui},
  \bibinfo{person}{Shu Wu}, \bibinfo{person}{Xiaoyu Zhang}, {and}
  \bibinfo{person}{Liang Wang}.} \bibinfo{year}{2019}\natexlab{}.
\newblock \showarticletitle{Semi-Supervised Compatibility Learning Across
  Categories for Clothing Matching}. In \bibinfo{booktitle}{\emph{2019 IEEE
  International Conference on Multimedia and Expo (ICME)}}. IEEE,
  \bibinfo{pages}{484--489}.
\newblock


\bibitem[\protect\citeauthoryear{Li, Ding, and Liu}{Li et~al\mbox{.}}{2018}]%
        {Zhongyang2018Constructing}
\bibfield{author}{\bibinfo{person}{Zhongyang Li}, \bibinfo{person}{Xiao Ding},
  {and} \bibinfo{person}{Ting Liu}.} \bibinfo{year}{2018}\natexlab{}.
\newblock \showarticletitle{Constructing Narrative Event Evolutionary Graph for
  Script Event Prediction}.
\newblock \bibinfo{journal}{\emph{arXiv preprint arXiv:1805.05081}}
  (\bibinfo{year}{2018}).
\newblock


\bibitem[\protect\citeauthoryear{Lian, Zhou, Zhang, Chen, Xie, and Sun}{Lian
  et~al\mbox{.}}{2018}]%
        {lian2018xdeepfm}
\bibfield{author}{\bibinfo{person}{Jianxun Lian}, \bibinfo{person}{Xiaohuan
  Zhou}, \bibinfo{person}{Fuzheng Zhang}, \bibinfo{person}{Zhongxia Chen},
  \bibinfo{person}{Xing Xie}, {and} \bibinfo{person}{Guangzhong Sun}.}
  \bibinfo{year}{2018}\natexlab{}.
\newblock \showarticletitle{xDeepFM: Combining explicit and implicit feature
  interactions for recommender systems}. In
  \bibinfo{booktitle}{\emph{Proceedings of the 24th ACM SIGKDD International
  Conference on Knowledge Discovery \& Data Mining}}. ACM,
  \bibinfo{pages}{1754--1763}.
\newblock


\bibitem[\protect\citeauthoryear{Liu, Yu, Wu, and Wang}{Liu
  et~al\mbox{.}}{2015}]%
        {liu2015convolutional}
\bibfield{author}{\bibinfo{person}{Qiang Liu}, \bibinfo{person}{Feng Yu},
  \bibinfo{person}{Shu Wu}, {and} \bibinfo{person}{Liang Wang}.}
  \bibinfo{year}{2015}\natexlab{}.
\newblock \showarticletitle{A convolutional click prediction model}. In
  \bibinfo{booktitle}{\emph{Proceedings of the 24th ACM international on
  conference on information and knowledge management}}. ACM,
  \bibinfo{pages}{1743--1746}.
\newblock


\bibitem[\protect\citeauthoryear{Marino, Salakhutdinov, and Gupta}{Marino
  et~al\mbox{.}}{2017}]%
        {marino2017more}
\bibfield{author}{\bibinfo{person}{Kenneth Marino}, \bibinfo{person}{Ruslan
  Salakhutdinov}, {and} \bibinfo{person}{Abhinav Gupta}.}
  \bibinfo{year}{2017}\natexlab{}.
\newblock \showarticletitle{The More You Know: Using Knowledge Graphs for Image
  Classification}. In \bibinfo{booktitle}{\emph{2017 IEEE Conference on
  Computer Vision and Pattern Recognition (CVPR)}}. IEEE,
  \bibinfo{pages}{20--28}.
\newblock


\bibitem[\protect\citeauthoryear{Mikolov, Sutskever, Chen, Corrado, and
  Dean}{Mikolov et~al\mbox{.}}{2013}]%
        {mikolov2013distributed}
\bibfield{author}{\bibinfo{person}{Tomas Mikolov}, \bibinfo{person}{Ilya
  Sutskever}, \bibinfo{person}{Kai Chen}, \bibinfo{person}{Greg~S Corrado},
  {and} \bibinfo{person}{Jeff Dean}.} \bibinfo{year}{2013}\natexlab{}.
\newblock \showarticletitle{Distributed representations of words and phrases
  and their compositionality}. In \bibinfo{booktitle}{\emph{Advances in neural
  information processing systems}}. \bibinfo{pages}{3111--3119}.
\newblock


\bibitem[\protect\citeauthoryear{Perozzi, Al-Rfou, and Skiena}{Perozzi
  et~al\mbox{.}}{2014}]%
        {perozzi2014deepwalk}
\bibfield{author}{\bibinfo{person}{Bryan Perozzi}, \bibinfo{person}{Rami
  Al-Rfou}, {and} \bibinfo{person}{Steven Skiena}.}
  \bibinfo{year}{2014}\natexlab{}.
\newblock \showarticletitle{Deepwalk: Online learning of social
  representations}. In \bibinfo{booktitle}{\emph{Proceedings of the 20th ACM
  SIGKDD international conference on Knowledge discovery and data mining}}.
  ACM, \bibinfo{pages}{701--710}.
\newblock


\bibitem[\protect\citeauthoryear{Qi, Liao, Jia, Fidler, and Urtasun}{Qi
  et~al\mbox{.}}{2017}]%
        {qi20173d}
\bibfield{author}{\bibinfo{person}{Xiaojuan Qi}, \bibinfo{person}{Renjie Liao},
  \bibinfo{person}{Jiaya Jia}, \bibinfo{person}{Sanja Fidler}, {and}
  \bibinfo{person}{Raquel Urtasun}.} \bibinfo{year}{2017}\natexlab{}.
\newblock \showarticletitle{3d graph neural networks for rgbd semantic
  segmentation}. In \bibinfo{booktitle}{\emph{Proceedings of the IEEE
  International Conference on Computer Vision}}. \bibinfo{pages}{5199--5208}.
\newblock


\bibitem[\protect\citeauthoryear{Qu, Cai, Ren, Zhang, Yu, Wen, and Wang}{Qu
  et~al\mbox{.}}{2016}]%
        {qu2016product}
\bibfield{author}{\bibinfo{person}{Yanru Qu}, \bibinfo{person}{Han Cai},
  \bibinfo{person}{Kan Ren}, \bibinfo{person}{Weinan Zhang},
  \bibinfo{person}{Yong Yu}, \bibinfo{person}{Ying Wen}, {and}
  \bibinfo{person}{Jun Wang}.} \bibinfo{year}{2016}\natexlab{}.
\newblock \showarticletitle{Product-based neural networks for user response
  prediction}. In \bibinfo{booktitle}{\emph{2016 IEEE 16th International
  Conference on Data Mining (ICDM)}}. IEEE, \bibinfo{pages}{1149--1154}.
\newblock


\bibitem[\protect\citeauthoryear{Qu, Fang, Zhang, Tang, Niu, Guo, Yu, and
  He}{Qu et~al\mbox{.}}{2018}]%
        {qu2018product}
\bibfield{author}{\bibinfo{person}{Yanru Qu}, \bibinfo{person}{Bohui Fang},
  \bibinfo{person}{Weinan Zhang}, \bibinfo{person}{Ruiming Tang},
  \bibinfo{person}{Minzhe Niu}, \bibinfo{person}{Huifeng Guo},
  \bibinfo{person}{Yong Yu}, {and} \bibinfo{person}{Xiuqiang He}.}
  \bibinfo{year}{2018}\natexlab{}.
\newblock \showarticletitle{Product-Based Neural Networks for User Response
  Prediction over Multi-Field Categorical Data}.
\newblock \bibinfo{journal}{\emph{ACM Transactions on Information Systems
  (TOIS)}} \bibinfo{volume}{37}, \bibinfo{number}{1} (\bibinfo{year}{2018}),
  \bibinfo{pages}{5}.
\newblock


\bibitem[\protect\citeauthoryear{Rendle}{Rendle}{2010}]%
        {rendle2010factorization}
\bibfield{author}{\bibinfo{person}{Steffen Rendle}.}
  \bibinfo{year}{2010}\natexlab{}.
\newblock \showarticletitle{Factorization machines}. In
  \bibinfo{booktitle}{\emph{2010 IEEE International Conference on Data
  Mining}}. IEEE, \bibinfo{pages}{995--1000}.
\newblock


\bibitem[\protect\citeauthoryear{Scarselli, Gori, Tsoi, Hagenbuchner, and
  Monfardini}{Scarselli et~al\mbox{.}}{2009}]%
        {scarselli2009graph}
\bibfield{author}{\bibinfo{person}{Franco Scarselli}, \bibinfo{person}{Marco
  Gori}, \bibinfo{person}{Ah~Chung Tsoi}, \bibinfo{person}{Markus
  Hagenbuchner}, {and} \bibinfo{person}{Gabriele Monfardini}.}
  \bibinfo{year}{2009}\natexlab{}.
\newblock \showarticletitle{The graph neural network model}.
\newblock \bibinfo{journal}{\emph{IEEE Transactions on Neural Networks}}
  \bibinfo{volume}{20}, \bibinfo{number}{1} (\bibinfo{year}{2009}),
  \bibinfo{pages}{61--80}.
\newblock


\bibitem[\protect\citeauthoryear{Shan, Hoens, Jiao, Wang, Yu, and Mao}{Shan
  et~al\mbox{.}}{2016}]%
        {shan2016deep}
\bibfield{author}{\bibinfo{person}{Ying Shan}, \bibinfo{person}{T~Ryan Hoens},
  \bibinfo{person}{Jian Jiao}, \bibinfo{person}{Haijing Wang},
  \bibinfo{person}{Dong Yu}, {and} \bibinfo{person}{JC Mao}.}
  \bibinfo{year}{2016}\natexlab{}.
\newblock \showarticletitle{Deep crossing: Web-scale modeling without manually
  crafted combinatorial features}. In \bibinfo{booktitle}{\emph{Proceedings of
  the 22nd ACM SIGKDD International Conference on Knowledge Discovery and Data
  Mining}}. ACM, \bibinfo{pages}{255--262}.
\newblock


\bibitem[\protect\citeauthoryear{Song, Shi, Xiao, Duan, Xu, Zhang, and
  Tang}{Song et~al\mbox{.}}{2018}]%
        {song2018autoint}
\bibfield{author}{\bibinfo{person}{Weiping Song}, \bibinfo{person}{Chence Shi},
  \bibinfo{person}{Zhiping Xiao}, \bibinfo{person}{Zhijian Duan},
  \bibinfo{person}{Yewen Xu}, \bibinfo{person}{Ming Zhang}, {and}
  \bibinfo{person}{Jian Tang}.} \bibinfo{year}{2018}\natexlab{}.
\newblock \showarticletitle{AutoInt: Automatic Feature Interaction Learning via
  Self-Attentive Neural Networks}.
\newblock \bibinfo{journal}{\emph{arXiv preprint arXiv:1810.11921}}
  (\bibinfo{year}{2018}).
\newblock


\bibitem[\protect\citeauthoryear{Tang, Qu, Wang, Zhang, Yan, and Mei}{Tang
  et~al\mbox{.}}{2015}]%
        {tang2015line}
\bibfield{author}{\bibinfo{person}{Jian Tang}, \bibinfo{person}{Meng Qu},
  \bibinfo{person}{Mingzhe Wang}, \bibinfo{person}{Ming Zhang},
  \bibinfo{person}{Jun Yan}, {and} \bibinfo{person}{Qiaozhu Mei}.}
  \bibinfo{year}{2015}\natexlab{}.
\newblock \showarticletitle{Line: Large-scale information network embedding}.
  In \bibinfo{booktitle}{\emph{Proceedings of the 24th international conference
  on world wide web}}. International World Wide Web Conferences Steering
  Committee, \bibinfo{pages}{1067--1077}.
\newblock


\bibitem[\protect\citeauthoryear{Tieleman and Hinton}{Tieleman and
  Hinton}{2012}]%
        {tieleman2012lecture}
\bibfield{author}{\bibinfo{person}{Tijmen Tieleman} {and}
  \bibinfo{person}{Geoffrey Hinton}.} \bibinfo{year}{2012}\natexlab{}.
\newblock \showarticletitle{Lecture 6.5-rmsprop: Divide the gradient by a
  running average of its recent magnitude}.
\newblock \bibinfo{journal}{\emph{COURSERA: Neural networks for machine
  learning}} \bibinfo{volume}{4}, \bibinfo{number}{2} (\bibinfo{year}{2012}),
  \bibinfo{pages}{26--31}.
\newblock


\bibitem[\protect\citeauthoryear{Vaswani, Shazeer, Parmar, Uszkoreit, Jones,
  Gomez, Kaiser, and Polosukhin}{Vaswani et~al\mbox{.}}{2017}]%
        {vaswani2017attention}
\bibfield{author}{\bibinfo{person}{Ashish Vaswani}, \bibinfo{person}{Noam
  Shazeer}, \bibinfo{person}{Niki Parmar}, \bibinfo{person}{Jakob Uszkoreit},
  \bibinfo{person}{Llion Jones}, \bibinfo{person}{Aidan~N Gomez},
  \bibinfo{person}{{\L}ukasz Kaiser}, {and} \bibinfo{person}{Illia
  Polosukhin}.} \bibinfo{year}{2017}\natexlab{}.
\newblock \showarticletitle{Attention is all you need}. In
  \bibinfo{booktitle}{\emph{Advances in neural information processing
  systems}}. \bibinfo{pages}{5998--6008}.
\newblock


\bibitem[\protect\citeauthoryear{Veli{\v{c}}kovi{\'c}, Cucurull, Casanova,
  Romero, Lio, and Bengio}{Veli{\v{c}}kovi{\'c} et~al\mbox{.}}{2017}]%
        {velivckovic2017graph}
\bibfield{author}{\bibinfo{person}{Petar Veli{\v{c}}kovi{\'c}},
  \bibinfo{person}{Guillem Cucurull}, \bibinfo{person}{Arantxa Casanova},
  \bibinfo{person}{Adriana Romero}, \bibinfo{person}{Pietro Lio}, {and}
  \bibinfo{person}{Yoshua Bengio}.} \bibinfo{year}{2017}\natexlab{}.
\newblock \showarticletitle{Graph attention networks}.
\newblock \bibinfo{journal}{\emph{arXiv preprint arXiv:1710.10903}}
  (\bibinfo{year}{2017}).
\newblock


\bibitem[\protect\citeauthoryear{Wang, Fu, Fu, and Wang}{Wang
  et~al\mbox{.}}{2017}]%
        {wang2017deep}
\bibfield{author}{\bibinfo{person}{Ruoxi Wang}, \bibinfo{person}{Bin Fu},
  \bibinfo{person}{Gang Fu}, {and} \bibinfo{person}{Mingliang Wang}.}
  \bibinfo{year}{2017}\natexlab{}.
\newblock \showarticletitle{Deep \& cross network for ad click predictions}. In
  \bibinfo{booktitle}{\emph{Proceedings of the ADKDD'17}}. ACM,
  \bibinfo{pages}{12}.
\newblock


\bibitem[\protect\citeauthoryear{Wu, Tang, Zhu, Xie, and Tan}{Wu
  et~al\mbox{.}}{2018}]%
        {Wu2018Session}
\bibfield{author}{\bibinfo{person}{Shu Wu}, \bibinfo{person}{Yuyuan Tang},
  \bibinfo{person}{Yanqiao Zhu}, \bibinfo{person}{Xing Xie}, {and}
  \bibinfo{person}{Tieniu Tan}.} \bibinfo{year}{2018}\natexlab{}.
\newblock \showarticletitle{Session-based Recommendation with Graph Neural
  Networks}. In \bibinfo{booktitle}{\emph{Thirty-Third AAAI Conference on
  Artificial Intelligence}}.
\newblock


\bibitem[\protect\citeauthoryear{Wu, Pan, Chen, Long, Zhang, and Yu}{Wu
  et~al\mbox{.}}{2019}]%
        {wu2019comprehensive}
\bibfield{author}{\bibinfo{person}{Zonghan Wu}, \bibinfo{person}{Shirui Pan},
  \bibinfo{person}{Fengwen Chen}, \bibinfo{person}{Guodong Long},
  \bibinfo{person}{Chengqi Zhang}, {and} \bibinfo{person}{Philip~S Yu}.}
  \bibinfo{year}{2019}\natexlab{}.
\newblock \showarticletitle{A comprehensive survey on graph neural networks}.
\newblock \bibinfo{journal}{\emph{arXiv preprint arXiv:1901.00596}}
  (\bibinfo{year}{2019}).
\newblock


\bibitem[\protect\citeauthoryear{Xiao, Ye, He, Zhang, Wu, and Chua}{Xiao
  et~al\mbox{.}}{2017}]%
        {xiao2017attentional}
\bibfield{author}{\bibinfo{person}{Jun Xiao}, \bibinfo{person}{Hao Ye},
  \bibinfo{person}{Xiangnan He}, \bibinfo{person}{Hanwang Zhang},
  \bibinfo{person}{Fei Wu}, {and} \bibinfo{person}{Tat-Seng Chua}.}
  \bibinfo{year}{2017}\natexlab{}.
\newblock \showarticletitle{Attentional factorization machines: Learning the
  weight of feature interactions via attention networks}.
\newblock \bibinfo{journal}{\emph{arXiv preprint arXiv:1708.04617}}
  (\bibinfo{year}{2017}).
\newblock


\bibitem[\protect\citeauthoryear{Zhang, Du, and Wang}{Zhang
  et~al\mbox{.}}{2016}]%
        {zhang2016deep}
\bibfield{author}{\bibinfo{person}{Weinan Zhang}, \bibinfo{person}{Tianming
  Du}, {and} \bibinfo{person}{Jun Wang}.} \bibinfo{year}{2016}\natexlab{}.
\newblock \showarticletitle{Deep Learning over Multi-field Categorical Data: A
  Case Study on User Response Prediction}.
\newblock \bibinfo{journal}{\emph{arXiv preprint arXiv:1601.02376}}
  (\bibinfo{year}{2016}).
\newblock


\bibitem[\protect\citeauthoryear{Zhou, Cui, Zhang, Yang, Liu, and Sun}{Zhou
  et~al\mbox{.}}{2018}]%
        {zhou2018graph}
\bibfield{author}{\bibinfo{person}{Jie Zhou}, \bibinfo{person}{Ganqu Cui},
  \bibinfo{person}{Zhengyan Zhang}, \bibinfo{person}{Cheng Yang},
  \bibinfo{person}{Zhiyuan Liu}, {and} \bibinfo{person}{Maosong Sun}.}
  \bibinfo{year}{2018}\natexlab{}.
\newblock \showarticletitle{Graph neural networks: A review of methods and
  applications}.
\newblock \bibinfo{journal}{\emph{arXiv preprint arXiv:1812.08434}}
  (\bibinfo{year}{2018}).
\newblock


\end{thebibliography}

% 
% If your work has an appendix, this is the place to put it.

\end{document}